\documentclass
[10pt,onecolumn,superscriptaddress,secnumarabic,amssymb,amsmath,nobibnotes,aps,prd,nofootinbib]{revtex4}
\usepackage[T1]{fontenc}
\usepackage{graphicx}
\usepackage{dcolumn}
\usepackage{bm,color}
\usepackage{hyperref}
\usepackage{accents}
\usepackage{amssymb,float}
\usepackage{amsmath}
\usepackage{multirow}
\usepackage{siunitx}
\usepackage{tabularx}
\usepackage{booktabs}
\usepackage{url}
\usepackage{orcidlink}
\usepackage{soul}
\usepackage{mathrsfs}

\usepackage{mathtools}

\DeclarePairedDelimiterX\braket[2]{\langle}{\rangle}{#1 \delimsize\vert #2}

\begin{document}

\title{Strengthening interacting agegraphic dark energy DGP constraints with local measurements and multimessenger forecastings}

\author{Maribel Hern\'andez}
\email{maribel.hernandez@correo.nucleares.unam.mx}
\affiliation{Instituto de Ciencias Nucleares, Universidad Nacional Aut\'{o}noma de M\'{e}xico, 
Circuito Exterior C.U., A.P. 70-543, M\'exico D.F. 04510, M\'{e}xico.}

\author{Celia Escamilla-Rivera\orcidlink{0000-0002-8929-250X}}
\email{celia.escamilla@nucleares.unam.mx}
\affiliation{Instituto de Ciencias Nucleares, Universidad Nacional Aut\'{o}noma de M\'{e}xico, 
Circuito Exterior C.U., A.P. 70-543, M\'exico D.F. 04510, M\'{e}xico.}

\begin{abstract}
An explanation of the nature of dark energy has been treated in extra dimensions within the scheme of string theory. One of the most successful models is inspired by the Dvali-Gabadadze-Porrati (DGP) model, in which the universe is a 4-dimensional brane embedded in a 5-dimensional Minkowski space-time. In this landscape, the study of the evolution of the normal branch
has led us to different kinds of dark energy, where the most simple case is the cosmological constant $\Lambda$. Moreover, other viable cosmological solutions are related to agegraphic dark energy,
which allows a late cosmic acceleration within an interacting mechanism. To explore the viability of these solutions and possible gravitational leakage, in this paper, we present constraints on such models using recent standard sirens forecasting in addition to local observables such as Pantheon (SNIa), $H(z)$ measurements, baryonic acoustic oscillations (BAO). Our results show that
the value associated with the species of quantum fields $n$ in these models is strongly restricted for supernovae observations to $n=20$, and for GW standard sirens mock data prefers a value of $n=1$.
\end{abstract}

\maketitle

\section{Introduction}

Since the discovery of the late time cosmic acceleration with measurements of Supernovae Type Ia (SNIa) \cite{riess1998observational,perlmutter1999measurements}, its explanation has been one of the most current intriguing issues in Cosmology. Furthermore, this phenomenon has been attributed to some kind of exotic component so-called \textit{dark energy}, which is characterized by negative pressure and on average constitutes the $70\%$ of content in the Universe. 

On this line of thought, the Cosmological Constant Cold Dark Matter ($\Lambda$CDM) model assumes that the cosmic acceleration is due to the existence of a Cosmological Constant $\Lambda$ which Equation-of-State (EoS) is $w_{\Lambda}=-1$. 
Although this model has been very successful and well constrained by local \cite{riess1998observational,abbott2022dark} and early \cite{aghanim2020planck} observables, it has several issues, e.g. the coincidence problem and the small value of $\Lambda$ \cite{Weinberg:2000yb}, and recently, cosmological tensions \cite{Abdalla:2022yfr}.
In particular, the $H_0$ tension \cite{DiValentino:2020zio} supports the idea that dark energy could be dynamical with an EoS $w_{\text{DE}}<-1$ \cite{Escamilla-Rivera:2021boq,Escamilla-Rivera:2019hqt,Zhang:2023zbj}. Among other proposals, dark energy can be associated with terms derived from alternative theories of gravity \cite{Jaime:2018ftn,Clifton:2011jh,Amendola:2006we}
and extended theories of gravity \cite{Bahamonde:2021gfp,Cai:2015emx}, showing a late cosmic acceleration.

Furthermore, different approaches have arisen to explain dark energy within the framework of fundamental theories such as quantum gravity \cite{Singh:2019hhi} or string theory \cite{Townsend:2021wrs}. Although, until today, there is still not a complete quantum gravity theory, some attempts have been made to discover the nature of dark energy using certain considerations, e.g. through the holographic principle \cite{Wang:2016och} we can propose the \textit{Holographic Dark Energy} (HDE) \cite{wang2017holographic} and employing the Károlyházy relation \cite{cai2007dark} we derive the \textit{Agegraphic Dark Energy} (ADE) \cite{cai2007dark}. 
In the case of the HDE model it has been found cosmological constraints with local observables \cite{Qiu:2021cww,Kim:2020cbm} and with Gravitational Waves (GW) from Einstein Telescope mock data \cite{Zhang:2019ple}. However, to obtain a cosmic acceleration solution, the length-scale considered in this model is the event horizon \cite{wei2009interacting}. 

In the case of the ADE model, the time scale is the age of the universe. Originally, this scale was pointed out as an advantage when compared to the HDE model \cite{cai2007dark}. However, it can be shown that the ADE solution never dominates the dynamics \cite{wei2008new}. To solve this drawback, it was proposed two different paths: \textit{(i)} considering an interaction between the ADE model and dark matter, so-called the ADE interacting model, which modifies the evolution of ADE in such a way that there is a dark energy stage-dominated phase \cite{wei2009interacting}.
\textit{(ii)} The second solution is the \textit{New Agegraphic Dark Energy} (NADE) \cite{wei2008new}, where the conformal time of the Friedmann-Robertson-Walker (FRW) universe is chosen to be the time scale instead of the age of the universe. Also, it has been considered an interaction scheme with dark matter for this model. The advantage of its description is that is a single-parameter model, unlike the two-parameter ADE model. Furthermore, this NADE model has been constrained with SNIa, Cosmic Microwave Background radiation (CMB), and Large Scale Structure (LSS) data finding better systematics in comparison to the $\Lambda$CDM model \cite{wei2008cosmological}.

In this line of thought, and to obtain general solutions that behave as dark energy, it has been proposed the existence of extra-dimensions within the framework of string theory, one of these is the so-called Dvali-Gabadadze-Porrati (DGP) model. In this scheme, the universe is a 4-dimensional brane embedded in a 5-dimensional Minkowski space-time \cite{deffayet2001cosmology} where gravity is modified at cosmological distances. Depending on how the brane is embedded into the space-time bulk, there are two kinds of cosmological solutions. In the well-known self-accelerating branch, we have a matter or radiation epoch that is followed by a late phase of accelerated expansion, while in the normal branch, there is not an accelerated phase and it is necessary to add some kind of dark energy to obtain it \cite{deffayet2001cosmology}. 
Nevertheless, the self-accelerating branch is disfavoured by SNIa and BAO \cite{fairbairn2006supernova} and this has led to focus the attention on the study of the evolution of the normal branch considering several cases, e.g. a $\Lambda$ \cite{sahni2003braneworld}, the HDE model \cite{wu2008dynamics}, the ADE and NADE models \cite{farajollahi2013cosmological}, and quintessence model \cite{chimento2003interacting}.
Specifically in \cite{ravanpak2019interacting} the evolution of the ADE model was studied within the framework of DGP brane-world cosmology considering an interaction between this ADE and dark matter. Using a dynamical system analysis, it is possible to find two critical points: an unstable point that is related to a matter-dominated era, and a second point related to an ADE-dominated era that is stable if the coupling parameter that determines the strength of interaction satisfies $\beta<1-2/(3n)$. This means that interaction could lead to a stable universe in the future, contrary to what happens in a DGP brane with non-interacting ADE, which does not have any stable critical point \cite{farajollahi2013cosmological}. 
Also, it was shown that there is no big-rip singularity in this model, in opposition to what happens in a DGP brane with HDE, which suffers from the big-rip singularity \cite{HDEDGP}. In this analysis, the cosmological parameters were constrained using independent measurements such as SNIa and Baryon Acoustic Oscillations (BAO) peaks, and it was found that these observations prefer a pure holographic dark energy or a pure DGP model \cite{HDEDGP}.

Furthermore, ADE and NADE models have been studied within the DGP braneworld scheme without considering interaction with dark matter, and its cosmological parameters have been constrained with $H(z)$ measurements, CMB, and BAO \cite{farajollahi2013cosmological}. In both scenarios, it was found that the universe undergoes an acceleration stage without entering the phantom regime, while in the matter-dominated effective dark energy vanishes. However, as far as we know, these models have not been studied with interaction terms that could lead to gravitational leakage effects. Therefore, in this paper, we studied these models by constraining them using current SNIa measurements (Pantheon (SN)) and standard sirens mock data \cite{Corman:2020pyr,Corman:2021avn} based on the Laser Interferometer Space Antenna (LISA) by forecasting multimessenger measurements of massive black hole binary (MBHB) mergers. Also, we will include in the analysis local measurements baselines as $H(z)$ measurements, and baryonic acoustic oscillations (BAO).

The paper is divided as follows:
In Sec.\ref{sec:theoryADE} we derive the evolution equations for the ADE and NADE models in a DGP braneworld with interactions. Also, we compute the deceleration parameter for each model.
In Sec.\ref{sec:observations} we describe the observables used to constrain the models described. We include SNIa Pantheon measurements, $H(z)$ measurements, baryonic acoustic oscillations (BAO) and, additionally, a GW mock catalog based on standard sirens. However, since we are dealing with extra-dimensional cosmological models, we need to write a new description of the luminosity distance $d_L$, in \ref{app:lum-theory} we derive this function for a D-dimensional space-time to obtain the luminosity gravity wave distance and also, we describe the GW data scattered for our DGP scenarios.
In Sec.\ref{sec:results} we discuss the statistical analysis,
and finally in Sec.\ref{sec:conclusions} we present our conclusions.


\section{Agegraphic models in DGP}
\label{sec:theoryADE}

\subsection{Agegraphic dark energy model (ADE).}
According to the DGP model, our universe is a brane embedded in a 5-dimensional Minkowski space-time. The consequence of this is related 
to a crossover scale $r_0$ that controls the transition between 4-dimensional behaviour to 5-dimensional, i.e. the gravitational potential of a source of mass $m$ is the well-known 4-dimensional gravitational potential $V\approx-Gm/r$ for $r\ll r_0$, and $V\approx-{G_{\text{bulk}}m}/{r^2}$ when $r \gg r_0$ \cite{dvali20004d}.
According to \cite{deffayet2001cosmology} the Friedmann equation in the DGP model is
\begin{equation}
\label{friedmann}
    H^2=\left(\sqrt{\frac{\rho}{3M_p^2}+\frac{1}{4r_0^2}}+\epsilon\frac{1}{2r_0}\right)^2,
\end{equation}
where $H=\dot{a}/a$ is the Hubble parameter, $\epsilon=\pm 1$, and their different values of $\epsilon$ correspond to two different embeddings of the brane into the bulk. If $\epsilon=1$ a phase of matter or radiation cosmology is followed by a late phase of accelerated expansion. This solution is known as the accelerated branch.  If $\epsilon=-1$  there is not an accelerated phase and it is necessary to add some kind of dark energy to obtain an accelerated expansion. This solution is known as the normal branch.
In this work, we consider the normal branch and assume that the total energy density on the brane is $\rho=\rho_{ \text{ADE}}+\rho_{\text{DM}}$, which contains ADE density $\rho_{\text{ADE}}$ and dark matter density $\rho_{\text{DM}}$. The Friedmann equation for this latter case is given by
\begin{equation}
\label{friedmann2}
    H=\sqrt{\frac{\rho_\text{ADE}}{3M_p^2}+\frac{\rho_{\text{DM}}}{3M_p^2}+\frac{1}{4r_0^2}}-\frac{1}{2r_0}.
\end{equation}
Based on the K\'arolyh\'azy relation \cite{maziashvili2007space}, which says that the distance $t$ cannot be known with better
accuracy than $\delta t=\gamma t_p^{2/3}t^{1/3}$, where $\gamma$ is a numerical factor of order unity \cite{cosmologicalmazias}, we can consider that in an over length scale, the space-time consists of cells of $\delta t^3 \approx t_p^2 t$. Therefore, a cell is the minimal detectable unit of space-time over $t$. Then, using the fact that this cell has a finite time $t$, due to the time energy uncertainty relation, there exists a minimal cell $\delta t^3$ whose energy cannot be smaller than $E_{\delta t^3}\gtrsim t^{-1}$. The energy density of the quantum fluctuations in Minkowski space-time is $\rho_{\text{ADE}}\sim E_{\delta t^3} / \delta t^3 \sim 1/ t_p^2 t^2$. If $t$ is the age of the universe, then the ADE density is given by
\begin{eqnarray}
\label{densityade}
\rho_{\text{ADE}}&=&\frac{3n^2M_p^2}{T^2}, \quad\quad \text{where} \quad\quad T=\int_0^a\frac{da}{Ha}, 
\end{eqnarray}
where $n^2$ is a numerical factor that was introduced to parameterize for example the species of quantum fields or the effect of curved space-time \cite{CAI2007228}.
Differentiating Eq.(\ref{densityade}) with respect to cosmic time $t$ we obtain
\begin{equation}
\label{dadedt}
\dot{\rho}_{\text{ADE}}=-2H\rho_{\text{ADE}}\frac{\sqrt{\Omega_{\text{ADE}}}}{n}.
\end{equation}
For this case, we are going to consider an interaction between dark matter and ADE. Since there is not a fundamental theory that tells us something about $Q$, previous works introduced specific forms. In our analysis, we study a specific interaction that is preferred by interacting HDE \cite{feng2016revisit}, and which has been already studied within the framework of DGP theory in \cite{ravanpakinteracting}. Then 
\begin{eqnarray}
\label{continuityinteractionADE}
    \dot{\rho}_\text{DM}+3H\rho_\text{DM}&=&Q, 
    \\
\dot{\rho}_{\text{ADE}}+3H(1+w_{\text{ADE}})~ \rho_{\text{ADE}}&=&-Q, 
\end{eqnarray}
where $Q$ is given by 
\begin{equation}
\label{interaction}
    Q=3\beta H\frac{\rho_{\text{ADE}}~ \rho_\text{DM}}{\rho_{\text{ADE}}+\rho_{\text{DM}}}.
\end{equation}
$\beta$ is a positive number, as $Q>0$ energy flows from ADE to a dark matter component. 
As it is standard, we define the critical density parameters as
\begin{equation}
\label{densityparameters}
    \Omega_{\text{DM}}=\frac{\rho_\text{DM}}{3H^2M_p^2}, \hspace{.1cm}   \Omega_{\text{ADE}}=\frac{\rho_{\text{ADE}}}{3H^2M_p^2}=\frac{n^2}{H^2T^2},     \hspace{.1cm}    \Omega_r=\frac{1}{4r_0^2H^2},   \hspace{.1cm}  \Omega_k=-\frac{k}{H^2a^2},
\end{equation}
and their corresponding current values, denoted by the subscript $0$, are given by 
\begin{equation}
\label{currentdensityparameters}
\Omega_{\text{0DM}}=\frac{\rho_{\text{0DM}}}{3H_{0}^2M_p^2},     \hspace{.1cm}    \Omega_{\text{0ADE}}=\frac{\rho_{\text{0ADE}}}{3H_0^2M_p^2}=\frac{n^2}{H_0^2T_0^2},    \hspace{.1cm} \Omega_{0r}=\frac{1}{4r_0^2H_0^2},     \hspace{.1cm}  \Omega_{0k}=-\frac{k}{H_0^2a_0^2}.
\end{equation}
Using Eq.(\ref{densityparameters}), we can rewrite Eq.(\ref{friedmann2}) as
\begin{equation}
\label{constrain}
    1=\Omega_{\text{DM}}+\Omega_{\text{ADE}}-2\sqrt{\Omega_r},
\end{equation}
and $\Omega_{\text{DE}}=\Omega_{\text{ADE}}-2\sqrt{\Omega_r}$ can be interpreted like an effective dark energy term. Defining $-2\sqrt{\Omega_r}$ as $\Omega_{\text{DGP}}=-2\sqrt{\Omega_r}$, we have 
\begin{equation}   \Omega_{\text{DE}}=\Omega_{\text{ADE}}-2\sqrt{\Omega_r}=\frac{\rho_{\text{ADE}}}{3M_p^2H^2}-\frac{1}{r_0H}=\Omega_{\text{ADE}}+\Omega_{\text{DGP}}.
\end{equation}
Combining Eqs.(\ref{dadedt})-(\ref{continuityinteractionADE}), the ADE EoS parameter is given by
\begin{equation}
\label{omegaade}
 w_{\text{ADE}}=-1+\frac{2}{3n}\sqrt{\Omega_{\text{ADE}}}-\frac{Q}{3H\rho_{\text{ADE}}}=-1+\frac{2}{3n}\sqrt{\Omega_{\text{ADE}}}-\frac{\beta~\Omega_\text{DM}}{\Omega_{\text{ADE}}+\Omega_{\text{DM}}}.
\end{equation}

The evolution of $\rho_\text{{ADE}}$ and $\rho_{\text{DM}}$ can be found by solving the following set of differential equations: 
\begin{eqnarray}
\label{setade1}    
\dot{\rho}_{\text{ADE}}&=&-2H\rho_{\text{ADE}}\frac{\Omega_{\text{ADE}}}{n}, \\
    \dot{\rho}_{\text{DM}}&=&-Q-3H\rho_\text{DM}.
\end{eqnarray}
In this work, we are going to constrain the model parameter set $\Theta_{\text{ADE}}: \{H_0,n,\Omega_{0r},\Omega_{\text{0ADE}},\beta\}$ with observations.
Furthermore, 
when there is no interaction between the components of the dark sector $w^{\text{eff}}_{\text{DM}}=w_{\text{DM}}=0$, $\rho_{\text{DM}}\propto (1+z)^3$, and $\rho_{\text{DM}}/\rho_{0c}=\Omega_{\text{0DM}}(1+z)^3$, the current critical density is $\rho_{0c}=3M_p^2H_0^2$. Otherwise, when there is an interaction 
$w^{\text{eff}}_{\text{DM}}=- Q/(3H\rho_{\text{DM}})$,
and the previous expressions for $\rho_\text{{DM}}$ and  $\rho_{\text{DM}}/\rho_{\text{0c}}$ are no longer satisfied. 

In order to fulfill $\rho_{\text{DM}}(z=0)=\rho_{\text{0DM}}$, we assume that  
\begin{equation}
\label{rhomrhoc}
    \frac{\rho_{\text{DM}}}{3M_p^2H_0^2}\equiv\Omega_{\text{0DM}}~f_A(z), \quad \quad \text{and}  \quad \quad   \frac{\rho_{\text{ADE}}}{3M_p^2H_0^2}=\frac{T_0^2}{T^2}\Omega_{\text{0ADE}}=\Omega_{\text{0ADE}}~g_A(z),
\end{equation}
where $f_A(z=0)=1=g_A(z=0)$ and $g_A(z):= T_0^2/T^2$.
Replacing Eq.(\ref{rhomrhoc}) in Eqs.(\ref{setade1})-(\ref{friedmann2}) we obtain 
\begin{eqnarray}
\label{setade2f}
    \frac{df_A}{dz}&=&3\left(\frac{f_A}{1+z}\right)-\frac{\beta~\Omega_{\text{0ADE}}~g_A f_A}{(1+z)(\Omega_{\text{0ADE}}~g_A+\Omega_{\text{0DM}}~f_A)}, \\
    \label{setade2g}
    \frac{dg_A}{dz}&=&\frac{2\sqrt{\Omega_{\text{0ADE}}}~g_A^{3/2}}{n(1+z)H} H_0, \\
    \label{setade2h}
    H&=&H_0
\left(\sqrt{\Omega_{\text{0DM}}~f_A(z)+\Omega_{\text{0ADE}}~g_A(z)+\Omega_{\text{0r}}}-\sqrt{\Omega_{\text{0r}}}\right).
\end{eqnarray}
We can solve numerically the latter equations using the initial condition $f_A(z=0)=g_A(z=0)=1$, and find the evolution for $\Omega_\text{DM}$, $\Omega_{\text{ADE}}$, $\Omega_r$, from the following equations
\begin{eqnarray}
\label{parametrosdensidad}
\Omega_{\text{DM}}=\Omega_{0\text{DM}}~f_A(z)\left(\frac{H_0}{H}\right)^2,  \hspace{.1cm} \Omega_{\text{ADE}}=\Omega_{\text{0ADE}}~g_A(z)\left(\frac{H_0}{H}\right)^2,\hspace{.1cm}\Omega_r=\Omega_{0r}\left(\frac{H_0}{H}\right)^2,\nonumber\\
\end{eqnarray}
and using Eqs.(\ref{parametrosdensidad})-(\ref{setade2h}) we can compute the deceleration parameter $q=-1-(\dot{H}/{H^2})$ as
\begin{eqnarray}
\label{decelerationade}
    q=-1+\frac{H}{H+H_0\sqrt{\Omega_{0r}}}\left(\frac{3}{2}-\frac{3}{2}\Omega_{\text{ADE}}+\frac{\Omega_{\text{ADE}}^{3/2}}{n}+3\frac{H_0}{H}\sqrt{\Omega_{0r}}+\frac{Q}{6M_p^2H^3}\right).\nonumber\\
\end{eqnarray}

\subsection{New Agegraphic dark energy model (NADE).}
In this model, the length scale is chosen to be the conformal time $\eta=\int_0^a da/Ha^2$ \cite{wei2008new},
then the energy density for the NADE model is 
\begin{equation}
\label{rhonade}
\rho_{\text{NADE}}=\frac{3n^2M_p^2}{\eta^2}, \quad\quad \text{and} \quad\quad \Omega_{\text{NADE}}=\frac{n^2}{H^2\eta^2}.
\end{equation}
Derivating Eq.(\ref{rhonade}) with respect to $t$ we obtain
\begin{equation}
\label{rhonadepunto}
\dot{\rho}_{\text{NADE}}=-2\rho_{\text{NADE}}~H\frac{\sqrt{\Omega_{\text{NADE}}}}{na},
\end{equation}
and considering an interaction between the NADE model and dark matter we have
\begin{eqnarray}
\label{continuitymatter}
\dot{\rho}_\text{DM}+3H\rho_{\text{DM}}&=&Q, \\
\dot{\rho}_{\text{NADE}}+3H(1+w_{\text{NADE}})~\rho_{\text{NADE}}&=&-Q, 
    \end{eqnarray}
where $Q$ is given by Eq.(\ref{interaction}).
Substituting Eq.(\ref{rhonadepunto}) in Eq.(\ref{continuitymatter}) we can find the EoS for the NADE model 
\begin{equation}
\label{omeganade}    
w_{\text{NADE}}=-1+\frac{2}{3na}\sqrt{\Omega_{\text{NADE}}}-\frac{Q}{3H\rho_{\text{NADE}}}=-1+\frac{2}{3na}\sqrt{\Omega_{\text{NADE}}}-\frac{\beta~\Omega_\text{DM}}{\Omega_{\text{NADE}}+\Omega_{\text{DM}}}.
\end{equation}
Finally, we have the following differential equations for $\rho_{\text{NADE}}$ and $\rho_\text{DM}$,
\begin{eqnarray}
\label{setnade}    
\dot{\rho}_\text{NADE}&=&-2H\rho_\text{{NADE}}\frac{\Omega_\text{{NADE}}}{n}, \\
    \dot{\rho}_\text{DM}&=&-Q-3H\rho_{\text{DM}}, \\
H&=&\sqrt{\frac{\rho_\text{NADE}}{3M_p^2}+\frac{\rho_\text{DM}}{3M_p^2}+\frac{1}{4r_0^2}}-\frac{1}{2r_0}.
\end{eqnarray}

We rewrite the latter equations in terms of the model parameters as:
\begin{equation}
    \frac{\rho_\text{DM}}{3M_p^2H_0^2} := \Omega_{\text{0DM}} ~f_N(z) \quad \quad \text{and}  \quad \quad
\frac{\rho_{\text{NADE}}}{3M_p^3H_0^2}=\Omega_{\text{0NADE}}~g_N(z),
\end{equation}
where $f_N(z=0)=1$, $g_{N}(z=0)=1$. We find 
\begin{eqnarray}
\label{setnade2f}
    \frac{df_N}{dz}&=&3\frac{f_N}{1+z}-\frac{\beta~ \Omega_{\text{0NADE}}~ g_Nf_N}{(1+z)(\Omega_{\text{0NADE}}~ g_N+\Omega_{\text{0DM}} ~ f_N)},\\
\label{setnade2g}
    \frac{dg_N}{dz}&=&\frac{2\sqrt{\Omega_{\text{0NADE}}} ~ g_N^{3/2}}{nH}H_0, \\
    \label{setnade2h}
      H&=&H_0 \left(\sqrt{\Omega_{\text{0DM}}f_N(z)+\Omega_{\text{0NADE}}~g_N(z)+\Omega_{\text{0r}}}-\sqrt{\Omega_{\text{0r}}}\right).
\end{eqnarray}

As in the ADE model, we can solve numerically the latter equations using the initial condition $f_N(z=0)=g_N(z=0)=1$, and find the evolution of $\Omega_\text{DM}$, $\Omega_{\text{NADE}}$, $\Omega_r$, from the following equations
\begin{eqnarray}
\Omega_{\text{DM}}=\Omega_{0\text{DM}}~ f_N(z)\left(\frac{H_0}{H}\right)^2,  \hspace{.05cm} \Omega_{\text{NADE}}=\Omega_{\text{0NADE}}~ g_N(z)\left(\frac{H_0}{H}\right)^2,\hspace{.05cm}\Omega_r=\Omega_{0r}\left(\frac{H_0}{H}\right)^2,\nonumber\\
\end{eqnarray}
and $q$ is given by 
\begin{eqnarray}
\label{decelerationNADE}
    q=-1+\frac{H}{H+H_0\sqrt{\Omega_{0r}}}\left(\frac{3}{2}-\frac{3}{2}\Omega_{\text{NADE}}+\frac{(1+z)}{n}~\Omega_{\text{NADE}}^{3/2}+3\frac{H_0}{H}\sqrt{\Omega_{0r}}+\frac{Q}{6M_p^2H^3}\right).\nonumber\\
\end{eqnarray}

\section{Local observational catalogs and forecasting baselines}
\label{sec:observations}

In this work, we use four different observational catalogs to constrain the cosmological parameters of the models described. Below we describe them briefly. 

\begin{itemize} 
\item\textbf{Pantheon supernovae sample (SNIa).}
This catalog contains the information of the apparent magnitude in the B band $m_\text{B}$ of 1048 supernova events for a redshift range $z=(0.01,2.26)$. The apparent magnitude $m_\text{B}$ is related to the observed distance modulus through 
\begin{equation}
\label{muobs}
    \mu_{\text{obs}}=m_\text{B}-M.
\end{equation}
For our analysis, we calibrate the absolute magnitude to obtain $M=-19.1$. 
The theoretical distance modulus $\mu$ is related to the luminosity distance, $d_L$, as follows:
\begin{equation}
\label{muteo}
\mu(z)=5\log\left[\frac{d_L(z)}{1\text{Mpc}}\right]+25,
\end{equation}
where 
\label{luminositydistance}    
$d_L=a_0c(1+z)\int_0^z dz/H,$
and
   $ d/dz \left(d_L/1+z\right)=c/H, $
with $a_0=1$.
Notice that $d_L$ is obtained considering that light can only travel along our 4-dimensional space-time and a spatially flat and expanding universe. This luminosity distance is also named $d_L^{\text{EM}}$ because is inferred from electromagnetic observations. If light could travel in extra dimensions the luminosity distance would be modified by the presence of these according to Eq.\ref{eq:dlapp}. To perform the statistical analysis, the best-fit parameters for a specific model, using the SN Pantheon catalog, can be calculated by maximising the logarithm of the likelihood function given by
\begin{equation}
\label{loglikelihoodsn}
 \ln \mathcal{L}_{\text{SN}}(\mu_{\text{obs}}(z_i)|z_i,\sigma_i,\Theta)=-\frac{1}{2}
\left(\chi_{\text{SN}}^2+\sum_{n=1}^N \ln(2\pi\sigma_i^2)\right),
\end{equation}
where $\Theta$ denotes the vector of free parameters of the model and 
\begin{equation}
\label{chiade}
\chi_{\text{SN}}^2=\sum_{i=1}^N \frac{(\mu_{obs}(z_i)-\mu(z_i;\Theta))^2}{\sigma_i^2},
\end{equation}
where $\sigma_i^2$ is the variance for each measurement, $N$ is the number of SNIa in the total sample.
Using the Bayes theorem we can obtain the joint posterior probability distribution, which is related to the likelihood using
\begin{equation}
    p_{\text{SN}}(\Theta|\mu_{\text{obs}},z,\sigma)=\frac{\pi(\Theta)\mathcal{L}_{\text{SN}}(\mu_{\text{obs}}(z_i)|z_i,\sigma_i,\Theta)}{{\varepsilon}},
\end{equation}
where $\pi(\Theta)$ is the prior probability distribution and $\varepsilon$ is the evidence given by 
\begin{equation}
    \varepsilon= \int d\theta \pi(\Theta)\mathcal{L}_{\text{SN}}(\mu_{\text{obs}}(z_i)|z_i,\sigma_i,\Theta).
\end{equation}
This is a normalisation constant that is not necessary to compute for the best-fit parameter values. As we can notice, to compute the posterior probability we need to know the prior and compute the likelihood given by Eq.(\ref{loglikelihoodsn}).\\

\item \textbf{$H(z)$ measurements}. These measurements can offer a tool to obtain constraints on the Hubble rate $H(z)$ at different redshifts $z$. To use this catalog, we consider the thirty-one data points reported in Ref.~\cite{Moresco:2016mzx,2012JCAP...08..006M,Moresco:2015cya,Moresco:2012jh}. This technique involves the use of spectroscopic dating methodologies on passively-evolving galaxies to compute the age difference between two galaxies at different $z$. When we measure the age difference given by $\Delta z / \Delta t$, we can compute $H(z) = -(1+z)^{-1} \Delta z/ \Delta t$. To perform the MCMC analysis, we consider $\chi^2_\mathrm{CC}$ to compute the agreement between the theoretical Hubble parameter values $H(z_i,\Theta)$, with DGP model parameters, i.e. labelled $\Theta$, and the measured Hubble data values $H_{\mathrm{obs}}(z_i)$, with an error of $\sigma_H(z_i)$. The $\chi^2_\mathrm{H(z)}$ can be computed through
\begin{equation}\label{eq:ccdata}
    \chi^2_\mathrm{H(z)} = \sum^{30}_{i=1} \frac{\left(H(z_i,\Theta) - H_{\mathrm{obs}}(z_i)\right)^2}{\sigma^2_H(z_i)} \,.
\end{equation}\\

\item \textbf{Baryonic acoustic oscillations (BAO)}. This catalog includes data from the Hubble parameter measurements and the corresponding comoving angular diameter at 
$z_{\mathrm{eff}} = {0.978,1.230,1.526,1.944}$ from SDSS-IV BOSS DR12\cite{Zhao:2018gvb}. For the BAO data mentioned, we calculated the comoving angular diameter distance that is related to the luminosity distance $d_L$ using: 

\begin{equation}
\label{DV}
    D_V(z)=\left[\frac{cz}{H(z)}\frac{d_L(z)^2}{(1+z)^2}\right].
\end{equation}

Notice that we require to adopt a fiducial value of $r_{s,\mathrm{fid}}(z_d)=147.78 \mathrm{Mpc}$ and the comoving sound horizon at the baryon drag epoch $r_s(z_d)=147.6$ \cite{Planck:2018vyg} Mpc. The statistics can be computed through the $\chi^2$:
\begin{equation}\label{eq:chibao}
    \chi^2_{\text{BAO}}(\Theta) = \Delta G(z_i,\Theta)^T C_{\text{BAO}}^{-1}\Delta G(z_i,\Theta),
\end{equation}
where $\Delta G(z_i,\Theta) = D_V(z_i,\Theta)\times\left(\frac{r_{s,fid}(z_d)}{r_s(z_d)}\right)-D_{V\text{obs}}(z_i)\times\left(\frac{r_{s,fid}(z_d)}{r_s(z_d)}\right)$ and $C_{\text{BAO}}$ is the corresponding covariance matrix for the BAO observations, and $\Theta$ is the same free parameter vector described above.
\\

\item \textbf{Gravitational Waves mock data.}
We use a mock catalog of standard sirens used previously in \cite{escamilla2022dynamical}, which consists of standard sirens mock data \cite{Corman:2020pyr,Corman:2021avn} based on the Laser Interferometer Space Antenna (LISA) by forecasting multimessenger measurements of massive black hole binary (MBHB) mergers.
This data was simulated assuming the $\Lambda$CDM model, and the best fits are given by the Pantheon catalog with $H_0=72.8$[km/s/Mpc] and $\Omega_m=0.285$. The mock redshift was generated with the normalised intrinsic merger rate $\dot{n}(z)$,  for a range $z$ $\epsilon$ $[0,2.3]$.
The siren distance $D_S(z)=d_L^{\text{GW}}$ that is the GW luminosity distance of the source is given by 
\begin{equation}
    D_S(z)=D_{S_{\Lambda \text{CDM}}}(z)+\mathcal{N}(0,\sigma),
\end{equation}
where $\mathcal{N}(0,\sigma)$ is the normal Gaussian probability distribution. The subscript $0$ denotes the mean value and $\sigma$ the standard deviation.
The forecasting of this catalog includes 1000 GW $d_L$ of simulated events with their respective redshifts and uncertainties. To employ the full database with the 
the SNIa sample, we compute the best-fit parameters of the model $\Theta$ by maximising the logarithm of the likelihood function given by 
\begin{equation}
 \label{loglikelihoodGW}
    \ln\mathcal{L}_{\text{GW}}(d_{L_\text{m}}^{\text{GW}}(z_i)|z_i,\sigma_i,\Theta)=-\frac{1}{2}\left(\chi_{\text{GW}}^2+\sum_{n=1}^{1000}\ln(2\pi\sigma_i^2)\right)
\end{equation}
where
\begin{equation}
\label{chiGW}
    \chi^2_{\text{GW}}= \sum_{i=1}^{1000}\frac{(d_L^{\text{GW}}(z_i,\Theta)-d_{L_{\text{m}}}^{\text{GW}}(z_i))^2}{\sigma_{i_{\text{m}}}^2}.
\end{equation}
$d_L^{\text{GW}}$ is the theoretical GW luminosity distance and $d_{L_\text{m}}^{\text{GW}}(z_i)$ is the GW luminosity distance obtained from the forecasting at redshift $z_i$. The posterior is given by
\begin{equation}
    p(\Theta|d_{L_\text{m}}^{\text{GW}},z,\sigma)\propto\pi(\Theta)\mathcal{L}_{\text{GW}}(d_{L_\text{m}}^{\text{GW}}
 (z_i)|z_i,\sigma_i,\Theta).
\end{equation}
Since within the DGP framework our 4-dimensional brane is embedded in a 5-dimensional Minkowski space-time and gravity propagates through this extra dimension, the gravitational wave distance, the distance measured from gravitational events, e.g. binary BH coalescence differs from the inferred electromagnetic luminosity distance $d_L^{\text{EM}}$. These quantities can be related using the relation Eq.(\ref{luminositygwdistanced}). Therefore, the $d_{L}^{\text{GW}}$ can be rewritten in terms of the DGP model parameters as 
    \begin{equation}
\label{luminositygwdistance5}
d_L^{\text{GW}}=d_L^{\text{EM}}\left[1+\left(2H_0\sqrt{\Omega_{0r}}\frac{d_L^{\text{EM}}}{c(1+z)}\right)^m\right]^{\frac{1}{2m}}.
\end{equation}
\end{itemize}

Finally, considering the combination of the four catalogs described gives the logarithm of the total likelihood function 
\begin{equation}
\ln\mathcal{L}_{\text{SN+GW+H(z)+BAO}}=\ln\mathcal{L}_{\text{SN}}+\ln\mathcal{L}_{\text{GW}}+\ln\mathcal{L}_{\text{H(z)}}+\ln\mathcal{L}_{\text{BAO}}.
\end{equation}
This is the function we are going to employ to constrain the ADE and NADE models with and without interactions.

\section{Cosmological constraint analysis}
\label{sec:results}

For our statistical analysis of ADE and NADE models in a DGP braneworld cosmology we use the SN Pantheon sample, the mock catalog computed for standard sirens, a 4-data points set from BAO measurements and $H(z)$ 31-data points catalog. 
Also, we performed a $\chi^2$-statistics using the combination of these catalogs to estimate the best-fit values of the parameters involved and their corresponding posteriors distributions.
In the case of SN observations, to obtain the $\ln\mathcal{L}_{\text{SN}}$ we need to compute the theoretical $\mu(z)$. First, we obtain the luminosity distance $d_L$. In the case of the ADE model, we solve numerically the set of Eqs.(\ref{setade2f})-(\ref{setade2g}), and finally Eq.(\ref{setade2h}). For the NADE model, we solve numerically Eqs. (\ref{setnade2f})-(\ref{setnade2g}). The theoretical $\mu(z)$ is obtained for each redshift $z$ of the sample using Eq.(\ref{muteo}). Finally, we compute the $\ln\mathcal{L}_{\text{SN}}$ through Eq.(\ref{loglikelihoodsn}) for each model. 
For the $H(z)$ measurements, we use directly Eq.(\ref{eq:ccdata}) with the catalog described. And for the BAO measurements, we compute 
$D_V(z)$ through Eq. (\ref{DV}), respectively, to finally calculate Eq.(\ref{eq:chibao}).
Finally, we compute the theoretical luminosity GW distance $d_L^{\text{GW}}$ for each $z_i$ of the mock catalog using Eq.(\ref{luminositygwdistance5}) and derive $\ln\mathcal{L}_{\text{GW}}$ using Eq.(\ref{loglikelihoodGW}).\\

\subsection{Results}

In this section, we discuss the constraints for the ADE and NADE models considering interacting terms and without them. The statistical analysis is performed using the observational samples described along with the mock data processed with GW configurations.

All statistical confidence levels (C.L) discussed below correspond to 1 and 2$\sigma$, respectively. We compute the posteriors of the different models performing a Markov-chain Monte Carlo analysis using the \texttt{emcee}
\footnote{\href{https://emcee.readthedocs.io/en/stable/}
{emcee.readthedocs.io}} code and we combine the marginalized distributions for each fractional density of the models using the \texttt{ChainConsumer} \footnote{\href{https://samreay.github.io/ChainConsumer/}{samreay.github.io/ChainConsumer}} package.

\begin{itemize}
\item \textbf{ADE without interaction.}
In this model $\beta=0$ in Eq.(\ref{setade2f}) and the parameters to constrain are $\Theta_{\text{ADE}}=\{\Omega_\text{{0ADE}},\Omega_{0r},n, H_0\}$ for SN observations and $\Theta_{\text{ADE}}=\{\Omega_\text{{0ADE}},\Omega_{0r},n, H_0,m\}$ for GW's observations.\\
To obtain the priors we consider that the current effective dark energy is $\Omega_{\text{0DE}}=\Omega_{\text{0ADE}}-2\sqrt{\Omega_{0r}}$ which satisfies $\Omega_{\text{0DM}}+\Omega_{\text{0DE}}=1$. If
$0<\Omega_{\text{0DE}}<1$, then $0<\Omega_{\text{0ADE}}-2\sqrt{\Omega_{0r}}<1$, and $0<2\sqrt{\Omega_{0r}}<\Omega_{\text{0ADE}}<1+2\sqrt{\Omega_{\text{0r}}}$, where if $r_0>H_0^{-1}$, therefore
\begin{equation}
\label{or2}
    0<\Omega_{0r}=\frac{1}{4r_0^2H_0^2}\leq 0.25\hspace{0.5cm} \text{and} \hspace{0.5cm} 0<\Omega_{\text{0ADE}}<2.
\end{equation}
Furthermore, $ \Omega_{\text{0ADE}}=n^2 / H_0^2T_0^2$, where $T_0=(\int_0^1 da/Ha)^{1/2}$ denotes the age of the universe. 
Following \cite{krauss2003age}, the age of the Universe lies between $11.2~ \text{Gy} < T_0 < 21 ~\text{Gy}$, and the best fit age is $T_0=13.14$ Gy, therefore if $66<H_0[\text{km/s/Mpc}]<74$ then $n\sim 1$. But considering this result $0<\Omega_{\text{0ADE}}<2, ~\text{and} ~ 0<\Omega_{0r}<0.25$, then
the auto-correlation time is less than $\tau/N$.
However, if we reduce the range of the priors for $0<\Omega_\text{{ADE}}<1$, and $0<\Omega_{0r}<0.001$, we obtain Gaussian posteriors distributions for $\Omega_{\text{0ADE}}$ and $H_0$, with the SN, GW and the combined sample SN+GW. The mean values obtained for the posteriors of these parameters are shown at the bottom of Table \ref{tab:ade-wint}. 
From this analysis, we obtain that the mean current effective dark energy is higher in comparison to the observations.
On the other hand according to \cite{farajollahi2013cosmological} $n>2$, then if we consider the range for the prior for $n$ as $0<n<20$ and priors shown at top Table \ref{tab:ade-wint} we get Gaussian C.L for $\Omega_{\text{0ADE}}$ and $H_0$, see Figure \ref{figure1}.
We found that according to the SN Pantheon sample, the most likely value of $n$ tends to its upper limit. This cutoff set a lower current effective dark energy value indicating an older cosmic age. Furthermore, when we combine the full baseline, SN+GW+BAO+H(z), we notice a variation less than 5\% on the $H_0$ bestfits and is even lower for the $m$ parameter $\sim$ 2\%.
This can be seen in Figure \ref{figure1}, where the most probable value for $n\sim 20$. For the GW forecast, the most likely value for $n\sim 1.34$. Although the effective dark energy agrees with observations, computing $T_0=n/(H_0\sqrt{\Omega_{\text{0ADE}}})$ with the mean values reported at top Table \ref{tab:ade-wint} we found that using SN, GW, and SN+GW data an older cosmic age of the order of hundreds of giga years (Gy). Notice that none of these values matches the estimated age of the Universe \cite{krauss2003age}. And the crossover scale is on the order of hundreds of Gigaparsecs (Gpc). Therefore, the effect of extra dimension appears at very large distances.

\renewcommand{\arraystretch}{2}
\begin{table}[h!]
    \centering    
    \begin{tabular}{|c|c|c|c|c|c|}
        \hline        
\textbf{Parameters} & \textbf{Priors}& \textbf{SN} &\textbf{GW}& \textbf{SN+GW}&\textbf{SN+GW+H(z)+BAO}\\
  \hline  
  $\Omega_{\text{0ADE}}$ & (0,1)&  $0.785\pm 0.027$&$0.799\pm 0.095$& $0.751\pm 0.025$&$0.755\pm0.026$\\
  \hline
  $\Omega_{\text{0r}}$&$(0,0.0025)$&$\left( 12.6\pm 8.5 \right) \times 10^{-4}$ & $\left( 12.6\pm 8.5 \right) \times 10^{-4}$
  &$\left( 10.3\pm 8.2 \right) \times 10^{-4}$&$\left(9.3\pm7.7\right)\times10^{-4}$ \\
  \hline
  $n$ & (0,20)& $13.5\pm 4.7$& $4.2\pm 3.0$ &$14.6\pm 4.0$&$13.3\pm 4.7$\\
  \hline
   $H_0$[km/s/Mpc] & (66,74)&$71.54\pm 0.23$&$68.9\pm 1.2$&$71.33\pm 0.20$&$71.38\pm0.2$\\
   \hline
 $ m$ &(0.1,2) &-  &$1.35\pm 0.43$ & $1.71\pm 0.22$&$1.73\pm0.2$\\
  \hline
  $\Omega_{\text{0DE}}$&- &$0.714^{+0.006}_{+0.003}$&$0.728^{+0.074}_{-0.064}$&$0.687^{+0.010}_{+0.003}$&$0.694_{+0.004}^{+0.010}$\\
  \hline
$T_0$[Gyr]  &-&$208.384^{+66.954}_{-69.703}$ &$66.722^{+39.559}_{-46.053}$  & $231.087^{+57.720}_{-59.967}$&$209.806^{+68.595}_{-55.257}$ \\
  \hline  $r_0$ [Gpc]  &-&$590.686^{+448.153}_{-135.690}$   & $613.319^{+480.915}_{-147.484}$   & $655.240^{+799.980}_{-167.692}$  &$689.085^{+976.907}_{-180.838}$\\
\hline\hline
$\Omega_{0\text{ADE}}$& $(0,1)$&$0.913\pm 0.031$&$0.828\pm0.077$&$0.812\pm0.020$ &$0.817\pm0.022$ 
\\
\hline
   $\Omega_{0r}$ &(0,0.001) &$(4.4\pm3.4)\times 10 ^{-4}$&$(5\pm3.4)\times 10^{-4}$&$(2.6\pm2.3)\times 10^{-4}$&$\left(3.5\pm3.0\right)\times 10^{-4}$\\
   \hline
   $n$ & (0,2)&$1.903\pm 0.077$&$1.42\pm0.28$&$1.947\pm0.042$&$1.97\pm0.042$\\
   \hline
  $H_0$[km/s/Mpc]&$(66,74)$&$70.82\pm0.21$&$68.32\pm 0.84$&$70.34\pm 0.18$&$70.33\pm0.17$\\
  \hline
  $m$ &(0.1,2)&- &$1.31\pm 0.46$&$1.67\pm0.24$&$1.69\pm0.24$\\
  \hline
  $\Omega_{0\text{DE}}$&-&$0.871^{+0.017}_{-0.009}$ & $0.783^{+0.015}_{+0.014}$& $0.795^{+0.35}_{-0.067}$&$0.779^{+0.008}_{+0.001}$\\ 
  \hline
$ T_0$[Gyr] &-&$27.514^{+0.556}_{-0.573}$    &$22.348^{+2.932}_{-3.274}$    &$30.053^{+0.199}_{-0.202}$&$30.320^{+0.164}_{-0.166}$  \\
\hline
$r_0$[Gpc]    &-&$1009.738^{+1114.606}_{-253.598}$&$981.879^{+775.461}_{-233.543}$&$1322.519^{+2580.862}_{-361.613}$&$1140.030^{+1883.515}_{-305.494}$ \\
\hline
    \end{tabular}    
          \caption{Constraints for the ADE model without interaction. The first column denotes the free parameters of the model, the second column describes the priors considered, and the third-fourth-fifth columns include the constraints for each parameter using SN Pantheon, GW mock data, and the total sample SN+GW, respectively. In the last column, we include the bestfits for the total baseline SN+GW+BAO+H(z). In the sub Table below we include a second option of priors for $\Omega_{0r}$ and $n$. Notice that $\Omega_{0\text{DE}}$ is a derivable parameter and $m$ is not a variable for SN data.}
    \label{tab:ade-wint}
\end{table}

\item \textbf{ADE with interaction.}
We set a flat prior for $0<\beta<1$ for this model. This selection is due to the existence of a stable solution for $\beta<1-2/3n$, and $n\geq1$ \cite{ravanpak2019interacting}. 
Furthermore, if $\beta>1$, the evolution of density parameters of dark matter and dark energy deviates from the $\Lambda$CDM model at 2$\sigma$. The priors considered for the other parameters are shown in Table \ref{tab:ade-int} and were chosen in such a way that the posterior distributions of $\Omega_{\text{0ADE}}$ and $H_0$ were Gaussian.
We found that GW prefers a greater value of the interacting term $\beta$ than SN observations, and from their corresponding posterior distributions, it can be seen that the most likely value of $\beta$ is close to zero for SN data, while for GW data is close to 1. However, the combined data indicates that the most likely value of $\beta$ is close to zero. From Tables \ref{tab:ade-wint}-\ref{tab:ade-int} it can be seen that the current effective dark energy $\Omega_{0\text{DE}}$ is lower in comparison to when there is no interacting term. As in the previous case, the posterior for $\Omega_{0r}$ is flat for SN and GW observations. In Table \ref{tab:ade-int}, it can be found that the crossover scale $r_0$
is on the order of hundreds of Gpc. Furthermore, when we combine the full baseline, SN+GW+BAO+H(z) the $m$ parameter deviates around 1\% in comparison to the sample SN+GW, also showing no significant change on $H_0$.

\renewcommand{\arraystretch}{2}
\begin{table}[h!]
    \centering    
    \begin{tabular}{|c|c|c|c|c|c|}
   \hline 
\textbf{Parameters}&\textbf{Prior}&\textbf{SN}& \textbf{GW} &\textbf{SN  + GW}&\textbf{SN+GW+H(z)+BAO}\\
  \hline
  $\Omega_{\text{0ADE}}$ & (0,1)& $0.763\pm 0.049$& $0.720\pm 0.093$& $0.721\pm 0.047$&$0.751\pm0.035$\\
  \hline
  $\Omega_{0r}$&$(0,0.005)$& $\left( 2.5\pm 1.7 \right) \times 10^{-3}$&$\left( 2.6\pm 1.7 \right) \times 10^{-3}$&$\left( 1.7\pm 1.4 \right) \times 10^{-3}$&$\left(1.5\pm1.3\right)\times 10^{-3}$\\
  \hline
   $n$ & (0,20)&$13.7\pm 4.6$   & $5.1\pm 3.9$ & $14.8\pm 3.9$&$13.9\pm4.4$\\
   \hline
  $H_0$[km/s/Mpc] &(66,74)&$71.49\pm 0.23$ &$68.9\pm 1.2$ &  $71.25\pm 0.20$&$71.34\pm0.2$\\
  \hline
  $\beta$&(0,1) &$0.44\pm 0.33$&  $0.52\pm 0.33$ &$0.38\pm 0.30$&$0.123\pm0.097$\\
  \hline
  $m$ &(0,2)& -&$1.38\pm 0.41$ & $1.74\pm 0.20$&$1.75\pm0.19$ \\ 
  \hline\hline
  $\Omega_{\text{0DE}}$&- & $0.663^{+0.019}_{-0.005}$&$0.618^{+.063}_{-0.051}$& $0.638^{+0.018}_{+0.001}$&$0.673^{+0.014}_{-0.158}$\\
  \hline 
  $T_0$[Gyr]&-&$214.649^{+62.932}_{-66.785}$ &$85.349^{+53.964}_{-63.447}$ & $239.345^{+52.850}_{-56.515}$&$219.977^{+62.320}_{-65.569}$ \\
  \hline
  $r_0$[Gpc]   &-& $419.639^{+324.579}_{-96.918}$ &$426.958^{+311.594}_{-100.641}$ &$510.601^{+708.294}_{-133.543}$&$542.890^{+948.056}_{-387.896}$\\
  \hline
    \end{tabular}
     \caption{Constraints for the ADE model with interaction. The first column denotes the free parameters of the model, the second column describes the priors considered, and the third-fourth-fifth columns include the constraints for each parameter using SN Pantheon, GW mock data, and the total sample SN+GW, respectively. In the last column, we include the bestfits for the total baseline SN+GW+BAO+H(z).
  In the sub Table below we include a second option of priors for $\Omega_{0r}$ and $n$. Notice that $\Omega_{0\text{DE}}$ is a derivable parameter and $m$ is not a variable for SN data.}
        \label{tab:ade-int}
\end{table}

\item \textbf{NADE without interaction.} 
Unlike the ADE model, in the NADE model, the parameter $n$ is not related to the age of the universe, so we consider the flat prior $0<n<20$. We found that to obtain a Gaussian posterior distribution for $\Omega_\text{{0ADE}}$ and $H_0$ we have to consider the flat priors $0<\Omega_{\text{0NADE}}<0.9$ and the ones shown in Table \ref{tab:nade-wint}.
For this model, it can be seen that for GW mock data the most likely value of $n$ is close to one while its mean value is $n=7.4$, GW prefers a smaller value for $H_0$ than SN.
Again, the crossover scale is on the order of hundreds of Gpc but this scale is larger than the values found at the top of Table \ref{tab:ade-wint}  and Table \ref{tab:ade-int}. Furthermore, when we combine the full baseline, SN+GW+BAO+H(z) we notice that there is not a significant change in the $H_0$ and $m$ values.

\renewcommand{\arraystretch}{2}
\begin{table}[h!]
    \centering    
    \begin{tabular}{|c|c|c|c|c|c|}
   \hline 
\textbf{Parameters}&\textbf{Prior}& \textbf{SN}& \textbf{GW}&\textbf{SN+GW}&\textbf{SN+GW+H(z)+BAO}\\
  \hline
  $\Omega_{\text{0NADE}}$&$(0,1)$ & $0.784\pm 0.026$ & $0.750\pm 0.061$ &$0.751\pm 0.023$& $0.758\pm.025$\\
  \hline
   $\Omega_{0r}$ &$(0,0.002)$& $\left( 10.1\pm 6.8 \right) \times 10^{-4}$ &$\left( 10.1\pm 6.8 \right) \times 10^{-4}$&$\left( 8.4\pm 6.6 \right) \times 10^{-4}$&$\left(7.9\pm6.4\right)\times 10^{-4}$\\
   \hline
   $n$&$(0,20)$  &$13.2\pm 4.9$ & $7.4\pm 5.2$ & $14.4\pm 4.1$&$12.4\pm5.2$ \\ 
   \hline
   $H_0$[km/s/Mpc]&$(66,74)$ & $71.56\pm 0.23$ & $69.84\pm 0.82$ & $71.34\pm 0.20$&$71.3\pm0.21$\\
   \hline
   $m$ & $(0,2)$ & - & $1.36\pm0.41$& $1.7\pm0.22$&$1.72\pm0.21$\\
   \hline\hline 
   $\Omega_{\text{0DE}}$& - & $0.720^{+0.007}_{+0.001}$ & $0.686^{+0.042}_{-.033}$ & $0.693^{+0.008}_{+0.003}$&$0.701^{+0.006}_{+0.005}$\\
   \hline
   $r_0$[Gpc]&-&$659.569^{+498.040}_{-151.311}$  &$675.812^{+520.540}_{-159.426}$  &$725.468^{+846.128}_{-184.095}$&$748.493^{+974.315}_{-193.795}$  \\
   \hline
    \end{tabular}
  \caption{Constraints for the NADE model without interaction. The first column denotes the free parameters of the model, the second column describes the priors considered, and the third-fourth-fifth columns include the constraints for each parameter using SN Pantheon, GW mock data, and the total sample SN+GW, respectively. In the last column, we include the bestfits for the total baseline SN+GW+BAO+H(z). Notice that $\Omega_{0\text{DE}}$ is a derivable parameter and $m$ is not a variable for SN data. }
    \label{tab:nade-wint}
\end{table}


\item \textbf{NADE with interaction.} 
In this case, to obtain Gaussian posteriors for $H_0$ and $\Omega_{\text{0ADE}}$, we set the flat priors shown in Table \ref{tab:nade-wint}. The posteriors of the parameters are shown in Figure \ref{figure1} at the bottom right side. This model prefers a lower value of $\beta$ than the one reported for the ADE model. The effective dark energy obtained is lower and the crossover scale is of the order of Gpc. Notice that this behaviour is almost the same when there is no interaction. The best fit for $n$ and $H_0$ using SN+GW data are the same for the non-interacting NADE model. Moreover, with GW data this model prefers a lower $n$ value in comparison to the constraints obtained from SN and SN+GW, separately. Finally, when we combine the full baseline, SN+GW+BAO+H(z), the deviations in comparison to the sample SN+GW, for the $H_0$ and $m$ parameters, are quite small $\sim$ 1\%.

\begin{figure}
\label{figure1}
\includegraphics[width=7.5cm]{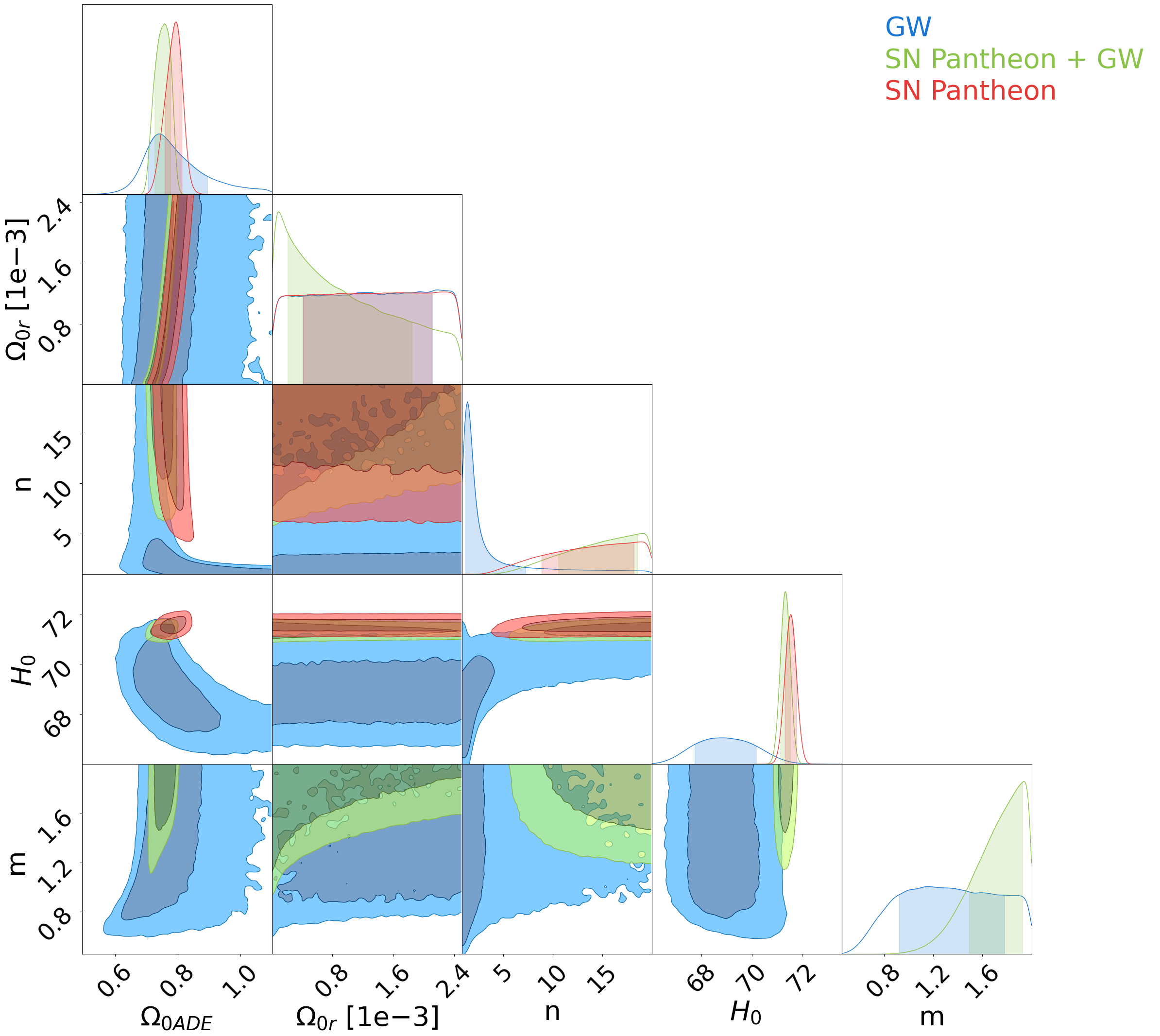}
\includegraphics[width=7.5cm]{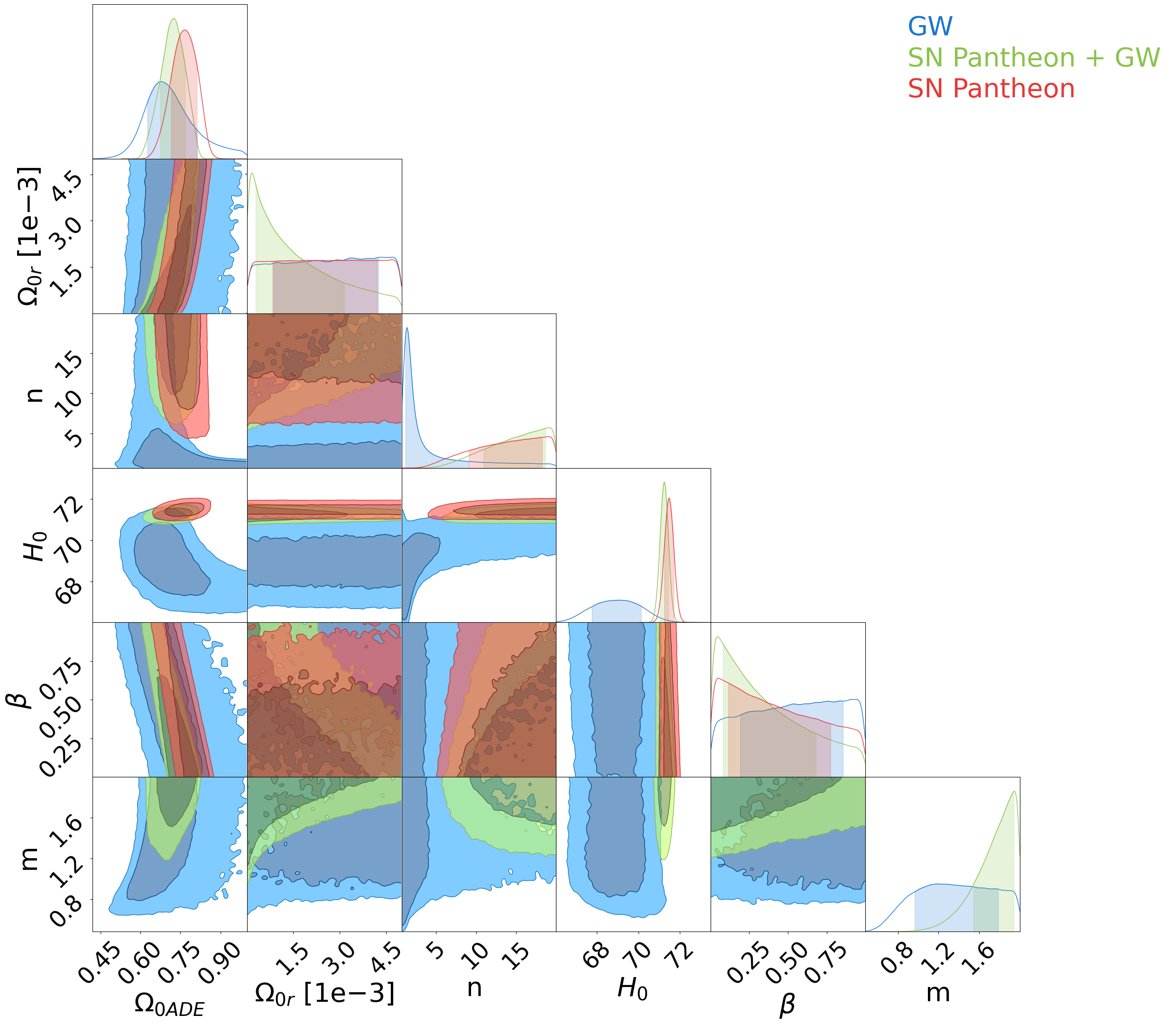}
\includegraphics[width=7.5cm]{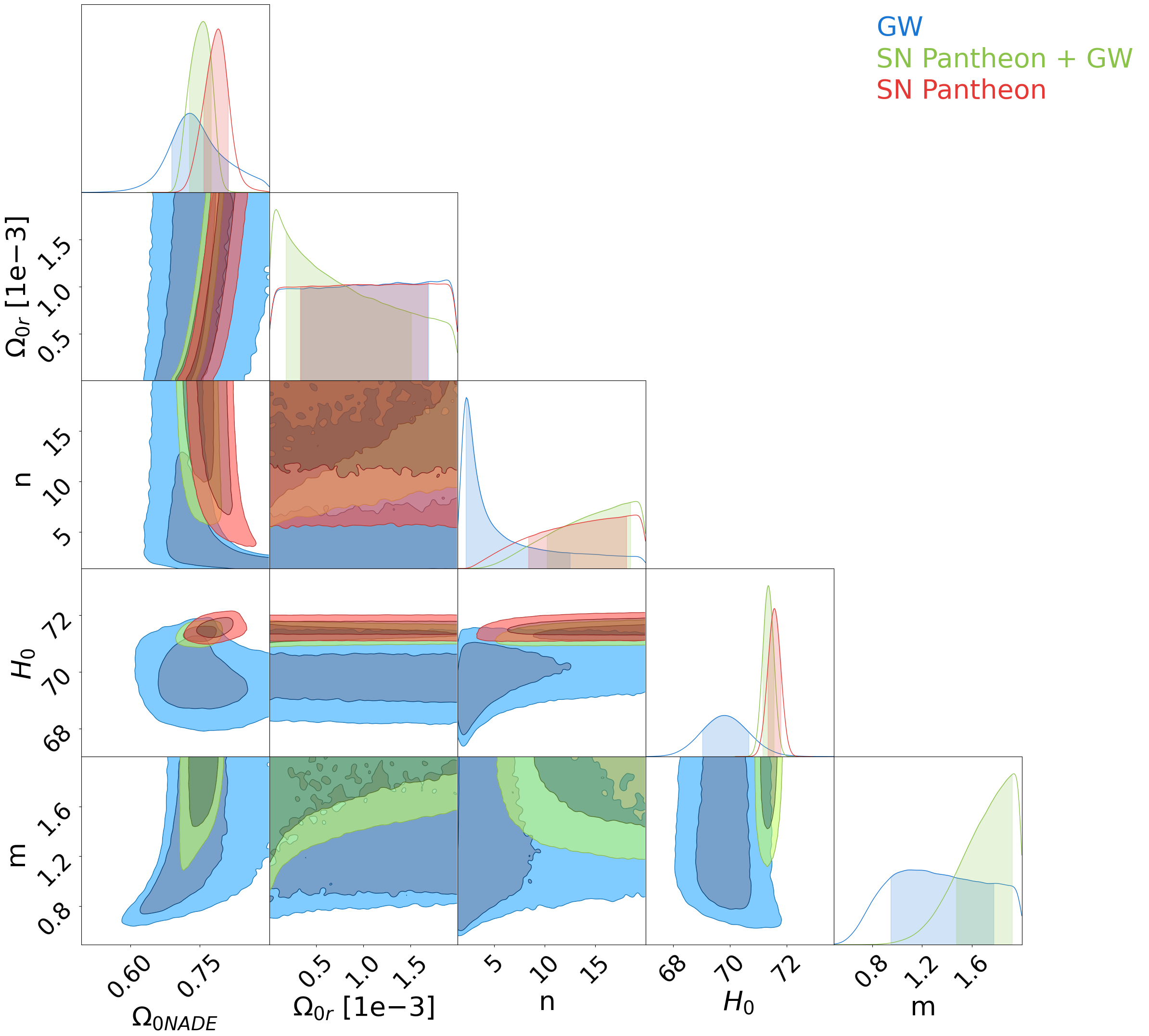}
\includegraphics[width=7.5cm]{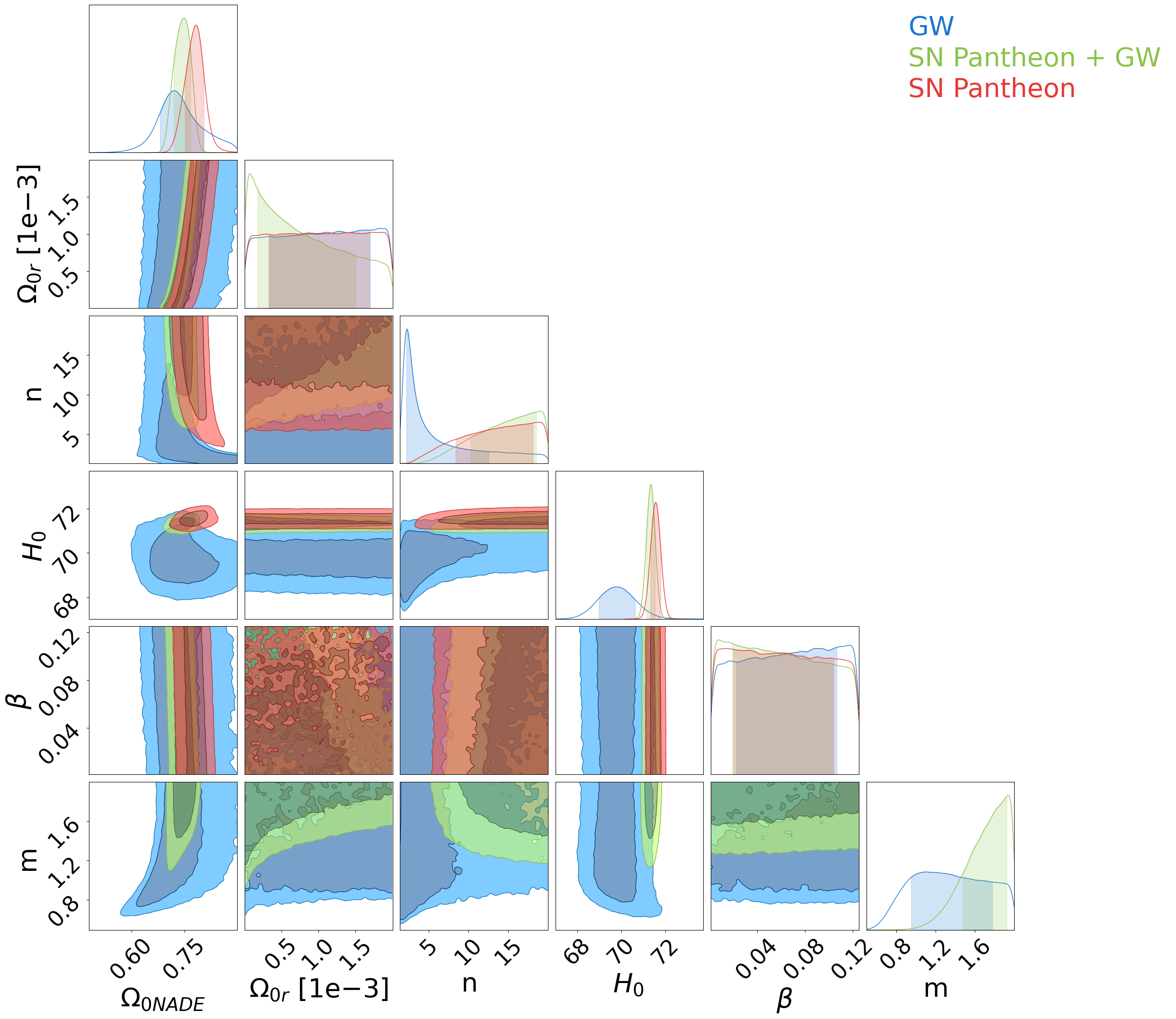}
\caption{2$\sigma$ C.L constraints using standard sirens mock data GW (blue), SN Pantheon (red), and the total sample SN Pantheon + GW (green) for the models discussed in Sec.\ref{sec:results}: \textit{Top left}: ADE without interaction. \textit{Top right}: ADE with interaction. \textit{Botton left}: NADE without interaction. \textit{Botton right}: NADE with interaction. }
\end{figure} 
\begin{figure}
\label{figure1}
\includegraphics[width=7.5cm]{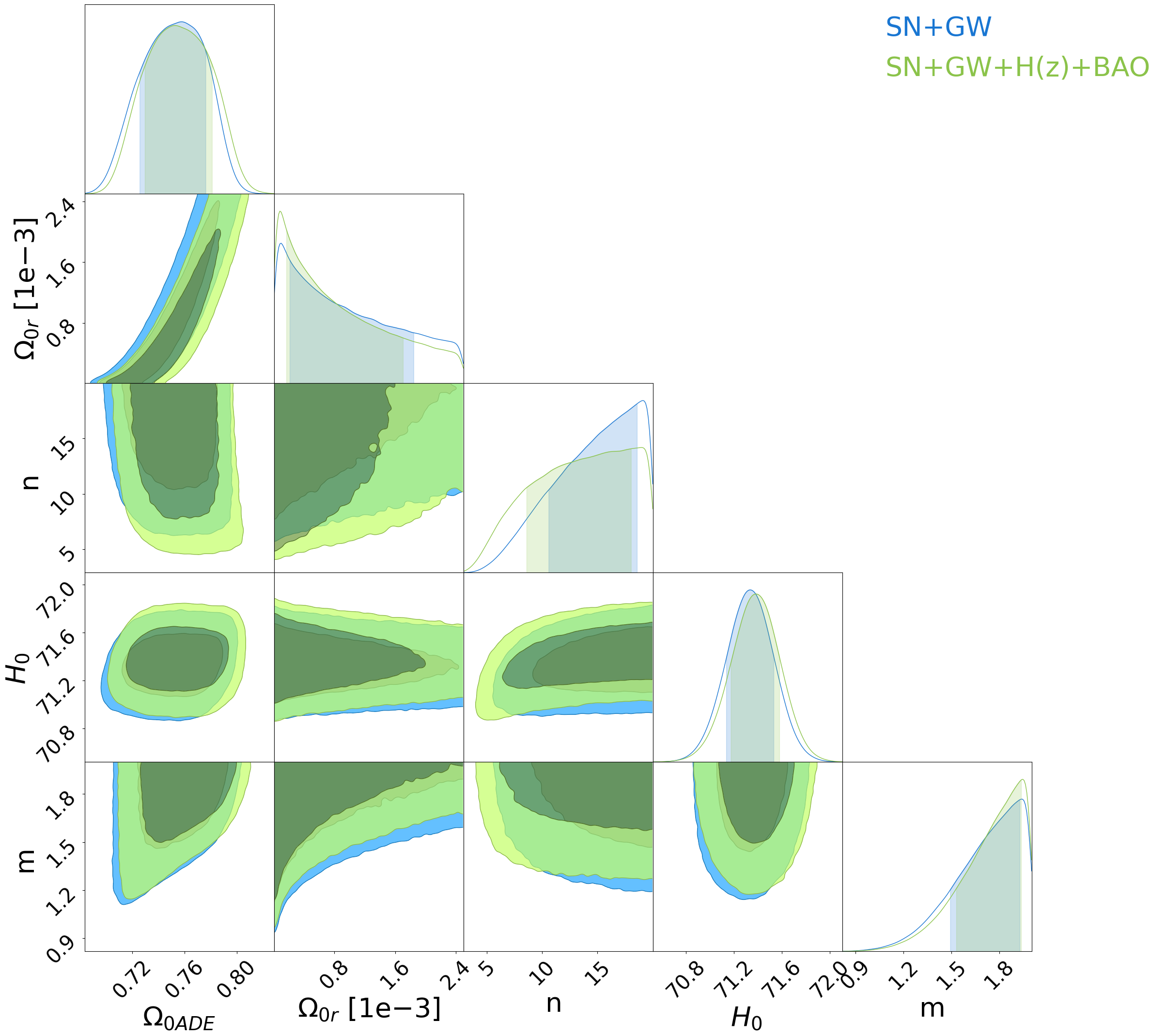}
\includegraphics[width=7.5cm]{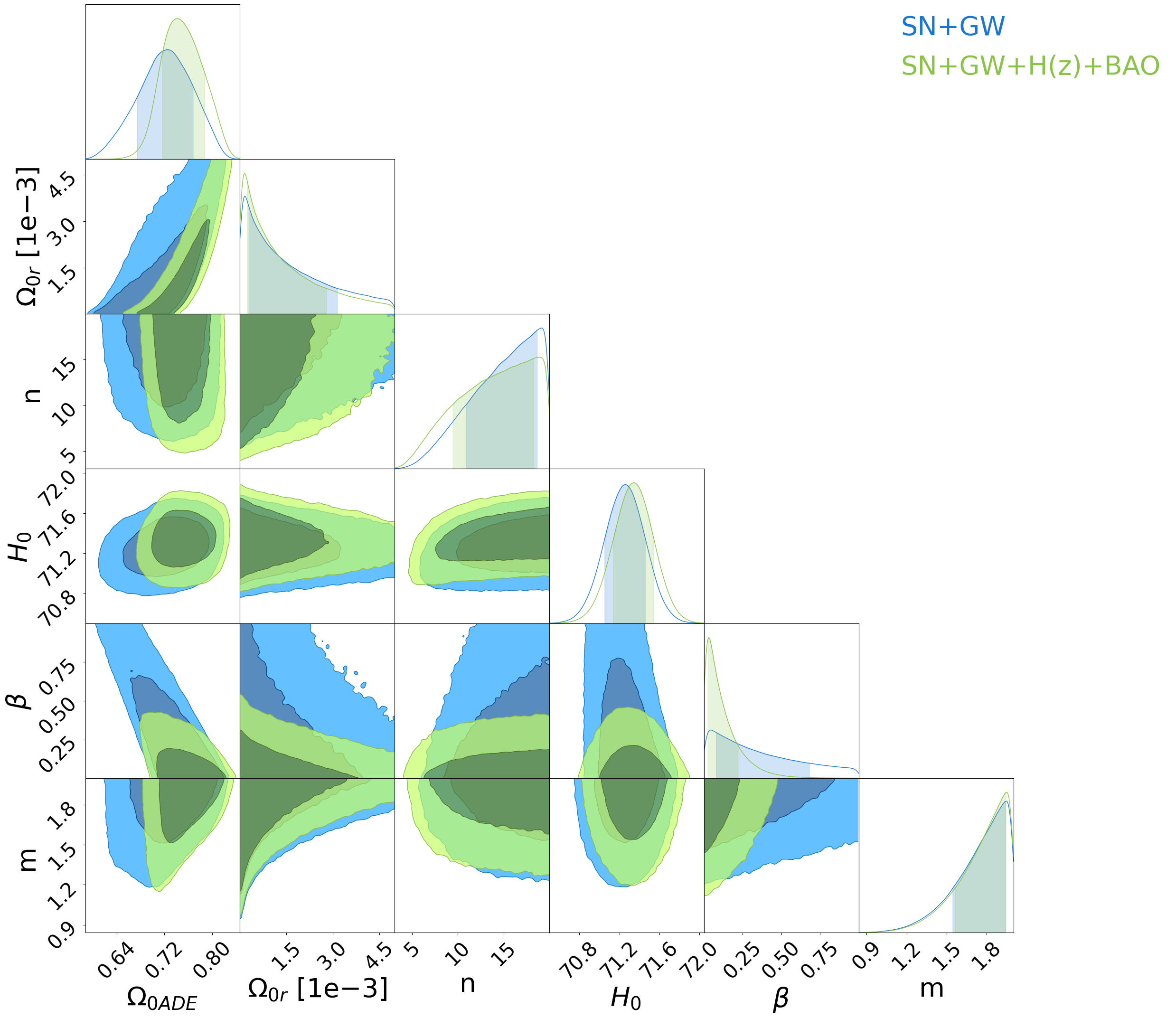}
\includegraphics[width=7.5cm]{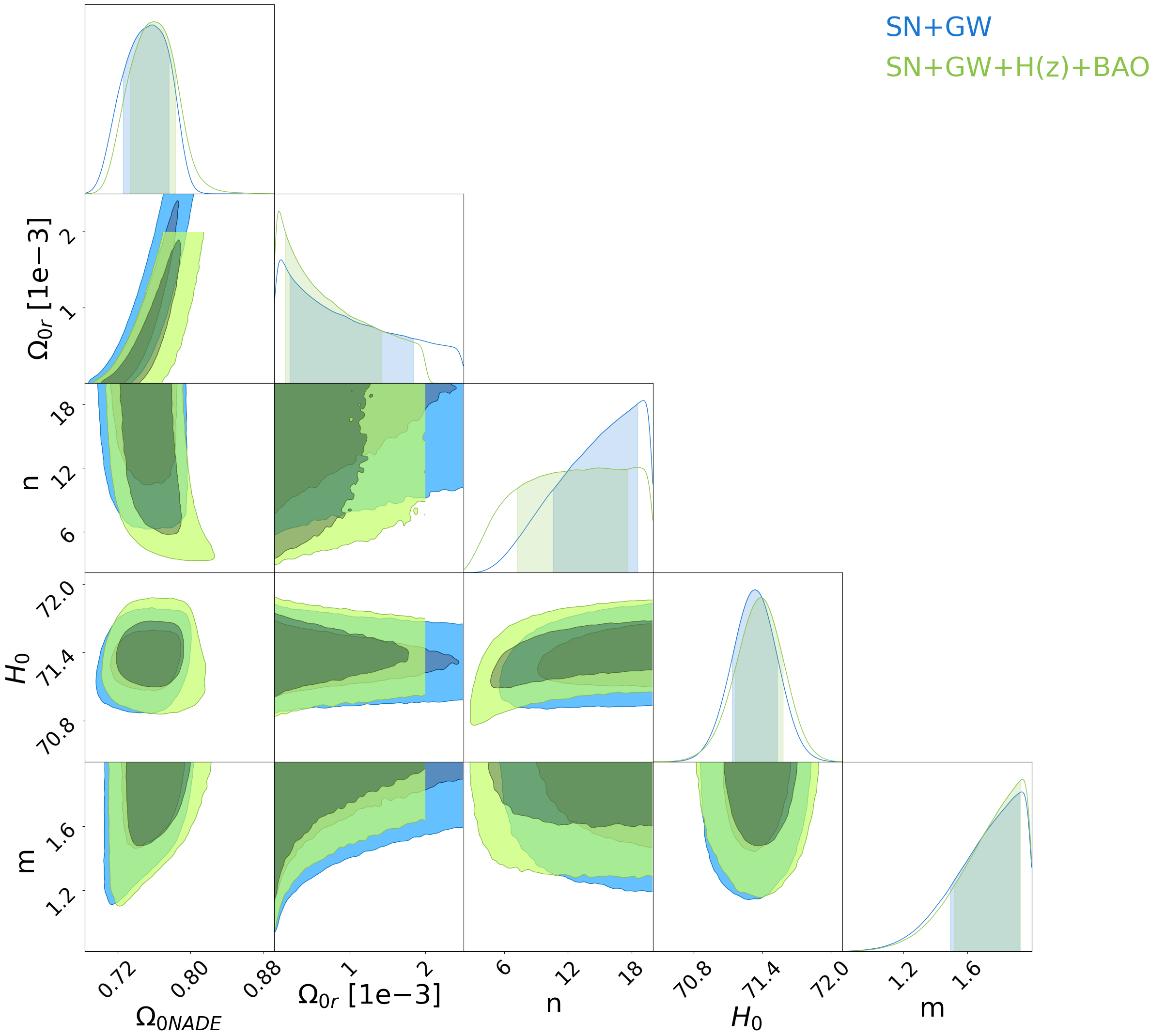}
\includegraphics[width=7.5cm]{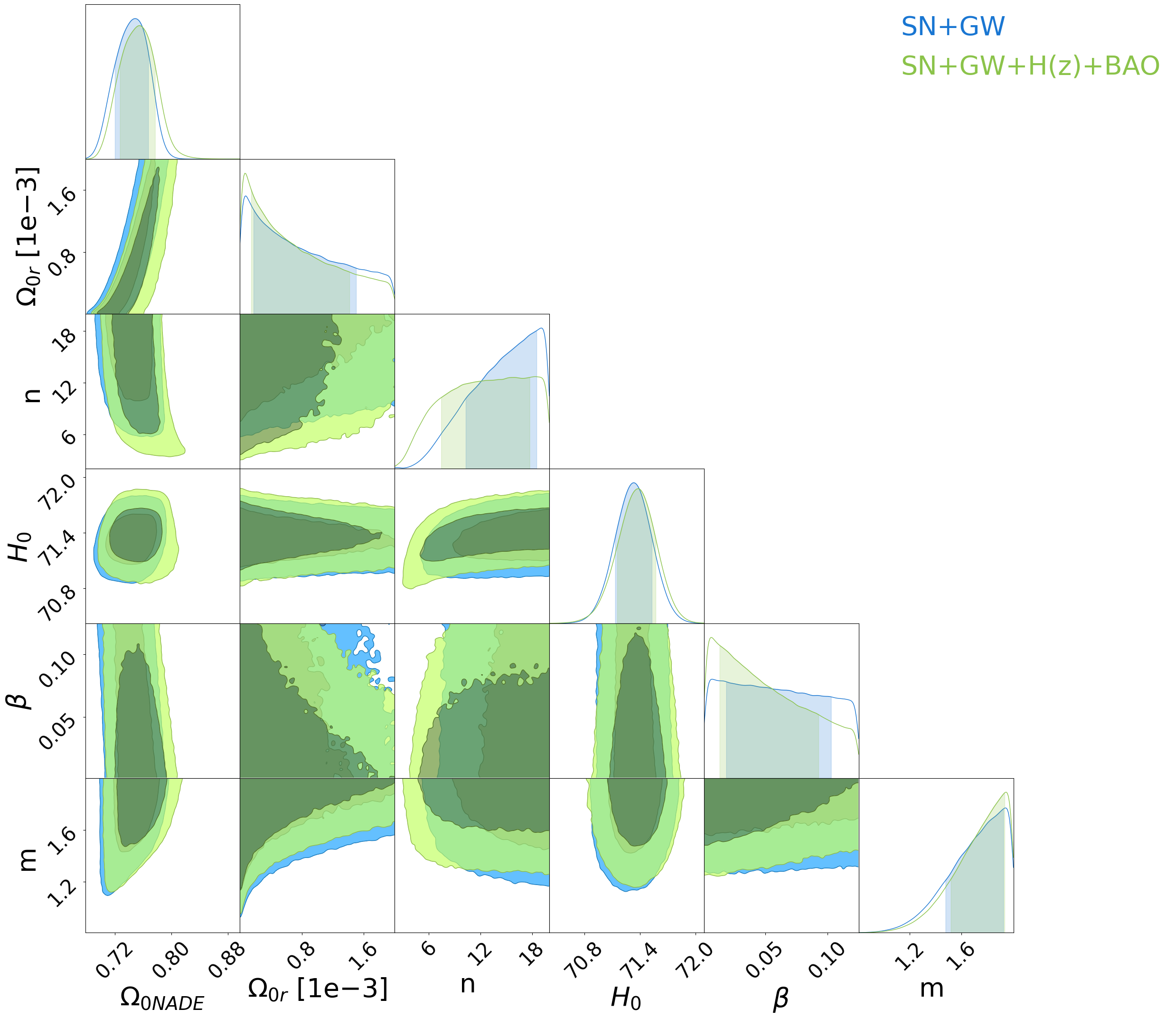}
\caption{2$\sigma$ C.L constraints using standard sirens mock data GW and SN Pantheon (blue), and the total sample SN+GW+H(z)+BAO (green) for the models discussed in Sec.\ref{sec:results}: \textit{Top left}: ADE without interaction. \textit{Top right}: ADE with interaction. \textit{Botton left}: NADE without interaction. \textit{Botton right}: NADE with interaction. }
\end{figure} 
\begin{figure*}
\label{evoluciondensidad}
\includegraphics[width=14cm]{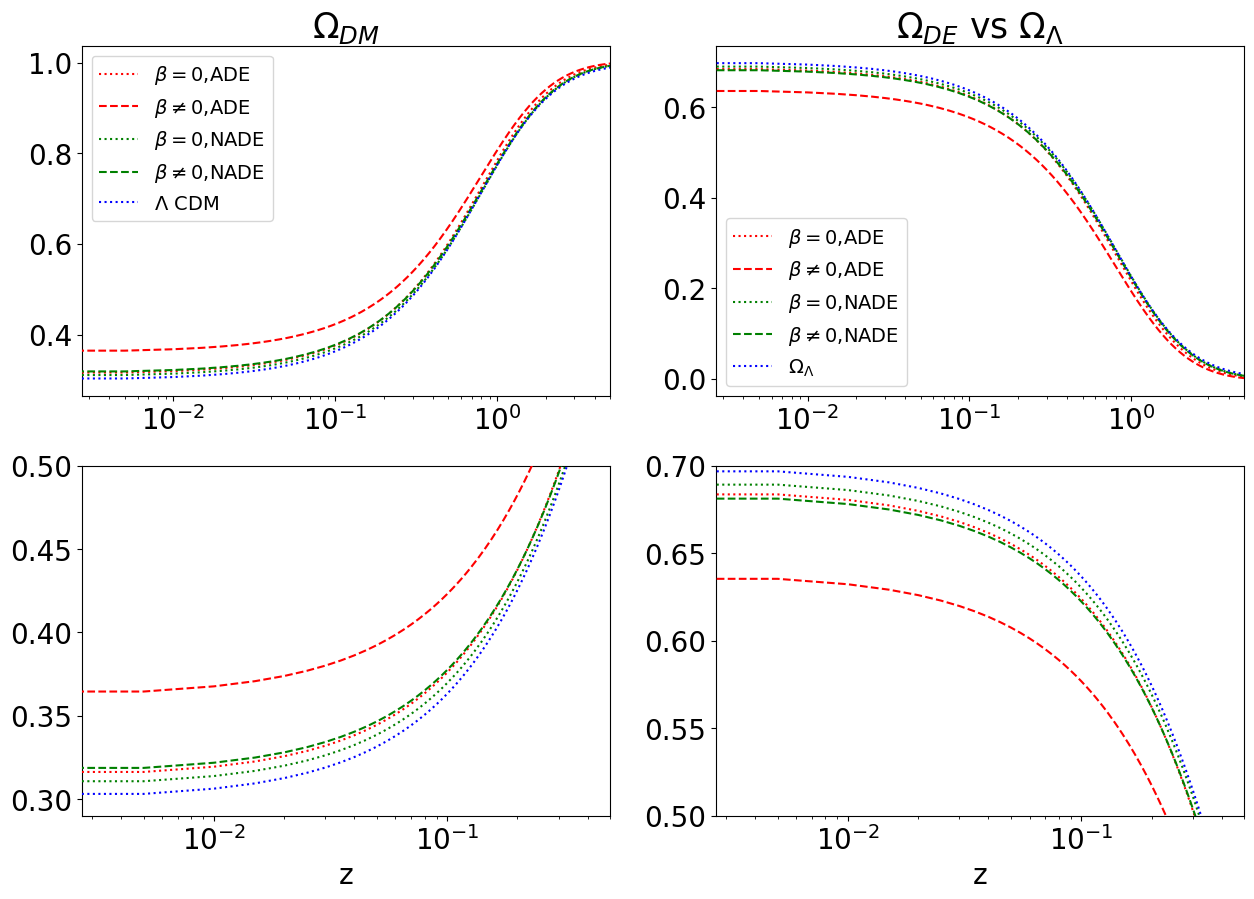}
\caption{Comparison of the evolution of density parameters with the $\Lambda$CDM model (dotted blue line) using the best-fit values of the parameters obtained from the SN+GW sample. The $\Lambda$CDM model is represented by the blue dotted line. For $z>5$  the evolution of the density parameters is the same as in the $\Lambda$CDM model. Each line denotes the models with and without interactions: ADE without interaction (dotted red line), ADE with interaction (dashed red line), NADE without interaction (dotted green line), and NADE with interaction (dashed green line).}
\end{figure*} 
\begin{figure*}
\label{evoluciondensidadhbao}
\includegraphics[width=14cm]{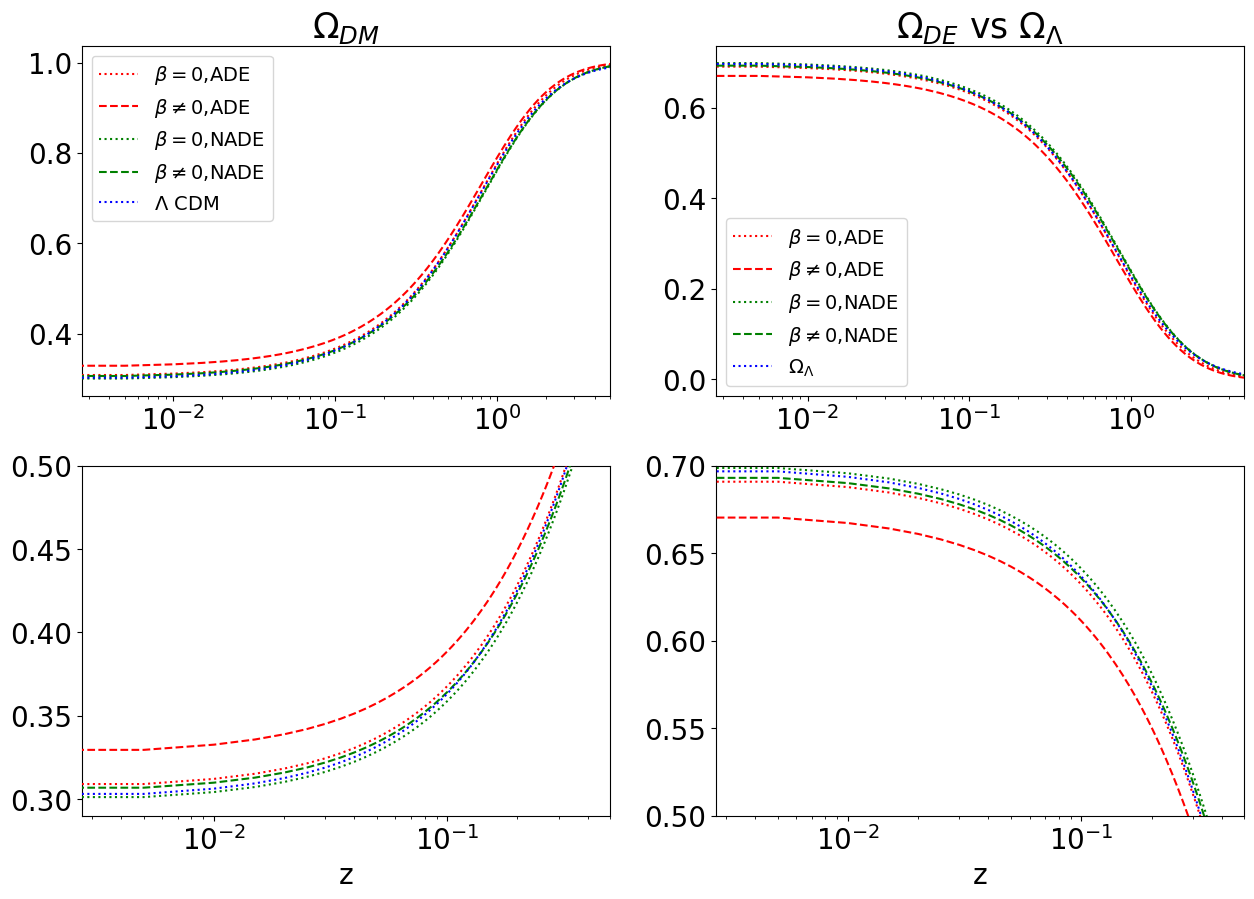}
\caption{Comparison of the evolution of density parameters with the $\Lambda$CDM model (dotted blue line) using the best-fit values of the parameters obtained from the SN+GW+H(z)+BAO sample. The $\Lambda$CDM model is represented by the blue dotted line. For $z>5$  the evolution of the density parameters is the same as in the $\Lambda$CDM model. Each line denotes the models with and without interactions: ADE without interaction (dotted red line), ADE with interaction (dashed red line), NADE without interaction (dotted green line), and NADE with interaction (dashed green line).}
\end{figure*}

\begin{figure*}
    \centering    
\label{parameterdeceleration}
    \centering
    \includegraphics[width=8.2cm]{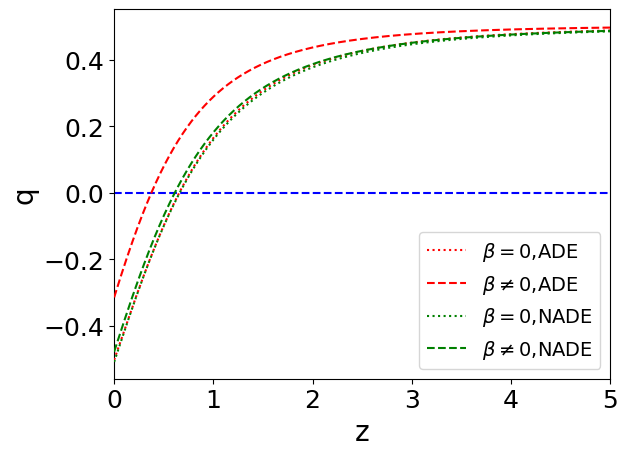}
    \caption{Deceleration parameter Eqs.(\ref{decelerationade}-\ref{decelerationNADE}) for ADE and NADE models using the best-fitted values obtained from the combined sample SN+GW. Each line denotes the models with and without interactions: ADE without interaction (dotted red line), ADE with interaction (dashed red line), NADE without interaction (dotted green line), and NADE with interaction (dashed green line). Each line denotes the models with and without interactions: ADE without interaction (dotted red line), ADE with interaction (dashed red line), NADE without interaction (dotted green line), and NADE with interaction (dashed green line).}
\end{figure*}
\begin{figure*}
    \centering  
\includegraphics[width=8.2cm]{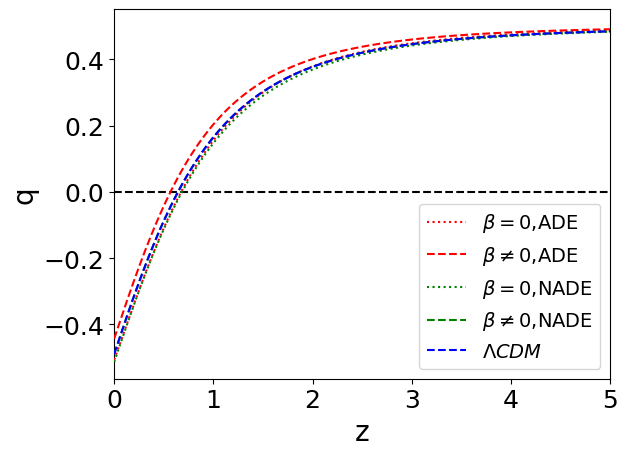}
    \caption{Deceleration parameter Eqs.(\ref{decelerationade}-\ref{decelerationNADE}) for ADE and NADE models using the best-fitted values obtained from the combined sample SN+GW+H(z)+BAO. Each line denotes the models with and without interactions: ADE without interaction (dotted red line), ADE with interaction (dashed red line), NADE without interaction (dotted green line), and NADE with interaction (dashed green line) and the $\Lambda$ CDM model is denoted by the dashed blue line.}
\end{figure*}

\renewcommand{\arraystretch}{2}
\begin{table}[h!]
    \centering    
    \begin{tabular}{|c|c|c|c|c|c|}
   \hline 
\textbf{Parameters} & \textbf{Prior}& \textbf{SN}&\textbf{GW} &\textbf{SN+GW}&\textbf{SN+GW+H(z)+BAO}\\
  \hline
  $\Omega_{\text{0NADE}}$&$(0,0.9)$ &$0.778\pm 0.027$ &$0.744\pm 0.062$&$0.743\pm 0.024$&$0.752\pm0.025$\\
  \hline
   $\Omega_{0r}$&$(0,0.002)$& $\left( 10.1\pm 6.8 \right) \times 10^{-4}$& $\left(10.1\pm 6.8 \right) \times 10^{-4}$ & $(8.4\pm6.6)\times 10^{-4}$&$\left(7.8\pm6.3\right)\times10^{-4}$\\
   \hline
    $n$ & $(0,20)$& $13.3\pm 4.8$& $7.3\pm 5.1$&$14.4\pm 4.1$&$12.6\pm5.1$\\
    \hline
    $H_0$[km/s/Mpc$]$ &$(66,74)$  & $71.55\pm 0.23$ &$69.80\pm 0.82$&$71.33\pm0.20$&$71.36\pm0.21$\\
    \hline
    $\beta$&$(0,0.125)$& $0.067\pm 0.039$&  $0.064\pm 0.042$&$0.061\pm0.042$&$0.053\pm0.040$\\
    \hline
    $m$ &$(0,2)$&- & $1.36\pm 0.41$& $1.70\pm 0.22$&$1.72\pm0.21$ \\
    \hline\hline 
    $\Omega_{\text{0DE}}$ &- & $0.714^{+0.008}_{-0.000}$&$0.680^{+0.043}_{-0.035}$&$0.685^{+0.007}_{+0.004}$&$0.696^{+0.006}_{-0.017}$\\
    \hline
    $r_0$[Gpc]&-&$659.661^{+498.110}_{-151.332}$ &$676.199^{+520.846}_{-159.521}$ &$725.570^{+846.247}_{-184.121}$&$752.643^{+968.712}_{-325.718}$\\
    \hline
    \end{tabular}
      \caption{Constraints for the NADE model with interaction. The first column denotes the free parameters of the model, the second column describes the priors considered, and the third-fourth-fifth columns include the constraints for each parameter using SN Pantheon, GW mock data, and the total sample SN+GW, respectively. In the last column, we include the bestfits for the total baseline SN+GW+BAO+H(z).
     Notice that $\Omega_{0\text{DE}}$ is a derivable parameter and $m$ is not a variable for SN data.}
    \label{tab:nade-int}
\end{table}
\end{itemize}

\section{Conclusions}
\label{sec:conclusions}

In this paper, we explore the observational constraints of interacting and non-interacting Agegraphic Dark Energy (ADE) and New Agegraphic Dark Energy (NADE) models in the normal branch within a DGP brane. Also, we include an analysis with the NADE to study the dynamics in the conformal time scheme instead of the age of the universe, as in ADE versions. A stable solution within these models is the one related to the evolution of the normal branch, which led us to different possibilities for dark energy. It is possible to recover the standard $\Lambda$CDM model in the case of a constant EoS, as we can see from each model in Eqs.\ref{omegaade}-\ref{omeganade}, respectively. Furthermore, viable cosmological solutions can be obtained from the ADE model version, which allows us to have a late cosmic acceleration within the interacting mechanism with dark matter.

In the ADE model the value of $n$ is related to the age of the universe, then if $T_0=13.4$ Gyr, and $66<H_0<74$, $n\approx 1$. However, when we considered in our analysis a prior within $0<n<2$, we found that neither the age of the universe nor the effective dark energy agrees with current observations (see Table \ref{tab:ade-wint}). Furthermore, when we extended the prior $n<20$, we found that the effective dark energy for the non-interacting and interacting ADE model agrees with observations, and the age of the universe is reduced when there are interactions. Conversely, the NADE model does not have the age drawback because $n$ does not depend on this cosmic age.
Furthermore, using the best-fitted values obtained for SN+GW (see Tables \ref{tab:ade-wint}-\ref{tab:ade-int}-\ref{tab:nade-wint}-\ref{tab:nade-int}), we obtain the density parameters evolution of the effective dark energy and dark matter for all models (see  Figure \ref{evoluciondensidad}). We found that the evolution of dark matter in these models deviates from the $\Lambda$CDM model at $z<5$. Nevertheless, in the interacting cases, there is more dark matter at present times. Also, we compute their respective deceleration parameters (see Figure \ref{parameterdeceleration}), from where we notice that the universe starts to accelerate at $z=0.655$ for both non-interacting ADE and NADE models. For the interacting ADE model starts to accelerate at $z\sim 0.37$, and for the interacting NADE model at $z\sim 0.61$. The reason for this behaviour is due to a bigger value of $\beta$. According to the constraints obtained using SN+GW data for all models, the best-fit values are $n\sim 14$ and $H_0\sim 71.3$km/s/Mpc. It is interesting to notice that the evolution of the density parameters in the non-interacting ADE and NADE models are approximately equal and that the evolution of their respective deceleration parameters is practically the same. We also notice that, generally, when we constraint using the total baseline SN+GW+BAO+H(z), there is not a significant change in the constraints reported for the $H_0$ and $m$ values. The effects of such BAO and H(z) catalogs should be related to the fact that they are calibrated with a concordance standard cosmological model. 

For all the models the value for $\Omega_{0r}$ is of the order of $10^{-4}$. In such a case,  $r_0$ cannot be constrained.   Furthermore, the value of $n$ is strongly restricted to an interval between $\{0,20\}$, where it was found that in the four models, SN observations prefer $n=20$ and GW mock data prefers a value of $n=1$. 
This result implies a younger universe in the ADE model with or without interaction in case we constrain them with GW than SN data.

For all models, GW data prefer a lower value for $H_0$ than SN data. The value of the current effective dark energy is less than the value for the case without interaction. In this analysis, we notice that the mean value of the crossover scale is larger, therefore the effects of the extra dimension appear at Gpc scales. At this scale, the dynamics of dark matter do not affect the cosmological evolution in ADE and NADE models.
\section*{Acknowledgments}
MH acknowledges financial support from SEP–CONAHCYT postgraduate grants program.
CE-R acknowledges funding from PAPIIT UNAM Project TA100122 and the Royal Astronomical Society as FRAS 10147. 
This research has been carried out using computational facilities procured through the Cosmostatistics National Group ICN UNAM project.
This article is based upon work from COST Action CA21136 Addressing observational tensions in cosmology with systematics and fundamental physics (CosmoVerse) supported by COST (European Cooperation in Science and Technology). 

\appendix

\section{Luminosity distance for gravitational waves in higher dimensions models}
\label{app:lum-theory}

In a Euclidean space, the energy per unit area or flux $F$, observed by a detector at a certain distance $D_L$ from a source of intrinsic luminosity $L$ is
   $ F= L / 4\pi d_L^2$.
Taking into account the cosmic expansion in a FRW universe, the luminosity observed is lowered by a factor $1/(1+z)^2$ and the flux is given by
\begin{equation}
\label{flux2}
    F=\frac{L}{4\pi a_0^2r^2(1+z)^2},
\end{equation}
where $a_0$  is the current value of the scale factor and $r$ is the comoving radial distance. From this relationship we establish 
the luminosity distance in a FRW universe without curvature is 
\begin{equation}
    d_L=(1+z)a_0r.
\end{equation}
Considering a $D=N+1$-dimensional space with a metric given by
\begin{equation}
\label{lineelement}
    ds^2=-dt^2+a(t)^2(dr_N^2+r_N^2d\Omega_{N-1}^2),
\end{equation}
where 
$d\Omega_{N-1}^2=d\theta_1^2+\sin^2\theta_1\sin^2\theta_2d\theta_3^2+...+\sin^2\theta_1...\sin^2\theta_{N-2}d\theta_{N-1}^2$, the Electromagnetic (EM) flux propagates isotropically in a hypersphere embedded in $N+1$ dimensions, implying that the relation now defines the luminosity distance in a D-dimensional space
\begin{equation}
\label{fluxd}
    F=\frac{L}{S_N},
\end{equation}
where $S_N$ is the area of a $N$-sphere of radius $r_N$ that can be obtained from integrating the line element at a fixed time and fixed radius $r_N$ as
\begin{eqnarray} 
\label{areahypersphere}
S_N&=&\int{d\theta_1}...d\theta_{N-1}\sqrt{-g_{N-1}}\nonumber\\
&=&\int{d\theta_1...d\theta_{N-1}r_N^{N-1}a(t)^{N-1}\times\sin^{N-2}\theta_1...\sin^2\theta_{N-2}\sin\theta_{N-1}}\nonumber\\
&=&b_{N-1}r_N^{N-1}a(t)^{N-1},
\end{eqnarray}
where
   $ b_{N-1}=2\pi^{N/2} / \Gamma(N/2). $
Therefore
\begin{equation}
\label{eq:ff}
    F=\frac{L}{b_{N-1}r_N^{N-1}a(t)^{N-1}}.
\end{equation}
Due to the cosmic expansion, the energy of each photon is also redshifted by 1/(1+z), this implies that the flux observed at Earth is lowered by a factor $1/(1+z)^2$, then Eq.(\ref{eq:ff}) is given by
\begin{equation}
    F=\frac{L}{b_{N-1}r_N^{N-1}a(t)^{N-1}(1+z)^2}=\frac{L}{b_{N-1}(d_L^D)^{N-1}},
\end{equation}
where $d_L^D$ denotes the luminosity distance for a $D$-dimensional space-time with metric (\ref{lineelement}).
From the previous equation
\begin{equation}
    d_L^D=a(t)r_N(1+z)^{2/(N-1)}=a(t)r_N(1+z)^{2/(D-2)}.
\end{equation}
Furthermore, if we consider in (\ref{lineelement})
$ds^2=0$ and radial geodesics, then 
\begin{equation}
    c^2dt^2=a(t)^2dr_N^2,
\end{equation}
and from this follows that 
\begin{equation}
    r_N= c \int_0^z\frac{dz}{H},
\end{equation}
then
\begin{equation}
    d_L^{(D)}=a_0c(1+z)^{\frac{2}{D-2}}\int_0^z\frac{dz}{H}.
\end{equation}
The lowest-order waveform of a binary system emitting GWs at cosmological distances in a 4-dimensional universe is 
\begin{equation}
    h_{\times}=\frac{4}{d_L^{4}}(G\mathcal{M}_{cz})^{5/3}(\pi \mathcal{f}_0)^{2/3}\cos\theta\sin\Phi(t_0),
\end{equation}
where $h_{\times}$ is the $\times$(cross)-polarization of the GW, $t_0$ and $f_0$ the time and frequency at the observer, $\mathcal{M}_{cz}$ is the redshifted chirp mass, $\theta$ the inclination angle, $\Phi$ the GW phase and 
\begin{equation}
    d_L^4=a_0(1+z)r_3,
\end{equation}
is the standard 4-dimensional luminosity distance 
\cite{maggiore2007gravitational}.
In the case of extra spatial dimensions, i,e. if we consider the metric given by \ref{lineelement}, the waveform of a binary system emitting GWs at cosmological distances in a D-dimensional space-time at the observer is \cite{Corman:2021avn}
\begin{equation}\label{eq:dlapp}
    h_\times\propto\frac{4}{(1+z)(a_0r_N)^{(D-2)/2}}(G\mathcal{M}_{cz})^{5/3}(\pi f_0)^{2/3}\cos\theta\sin\Phi(t_0),
\end{equation}
and we can rewrite this waveform as 
\begin{equation}
    h_\times\propto\frac{4}{d_L^{\text{GW}}}(G\mathcal{M}_{cz})^{5/3}(\pi f_0)^{2/3}\cos\theta\sin\Phi(t_0).
\end{equation}
We defined in the latter equation:
\begin{equation}
    d_L^{\text{GW}}\propto(1+z)(a_0r_N)^{(D-2)/2},
\end{equation}
with
$    d_L^{GW}\propto(d_L^D)^\frac{D-2}{2}$.
In the case of $r_N=r_3$, we can rewrite the above equation as
\begin{equation}
\label{luminosity4d}
    d_L^D=a_0r_3(1+z)^{2/(D-2)}=d_L^4(1+z)^{(4-D)/(D-2)},
\end{equation}
where $d_L^4$ is the luminosity distance that we can infer from observations denoted by $d_L^{\text{EM}}$, which is the luminosity distance from the electromagnetic signal emitted by a source. 
Combining the latter equations we obtain
\begin{equation}
    d_L^{\text{GW}}\propto d_L^{\text{EM}}\left(\frac{d_L^{\text{EM}}}{1+z}\right)^{\frac{D-4}{2}}.
\end{equation} 
In the DGP theory, we have a crossover scale $r_0$ such that if $r\ll r_0$, we recover 4D gravity, and if $r\gg r_0$ the effect of extra-dimensions appears as we can notice for example from the Friedmann equation. Then, to obtain a valid relation at all scales and recover 4-D gravity for $r\ll r_0$ we write 
\begin{equation}
\label{luminositygwdistanced}    d_L^{\text{GW}}=d_L^{\text{EM}}\left[1+\left(\frac{d_L^{EM}}{c r_0(1+z)}\right)^m\right]^\frac{D-4}{2m},
\end{equation}
here $m$ determines the steepness of the transition from the small-scale to large-scale behaviour \cite{Corman:2021avn}. The value of $m$ is a free parameter and has to be determined by observations, it has to be different from zero.

\begin{figure}[ht]
  \centering
  \includegraphics[width=0.49\textwidth]{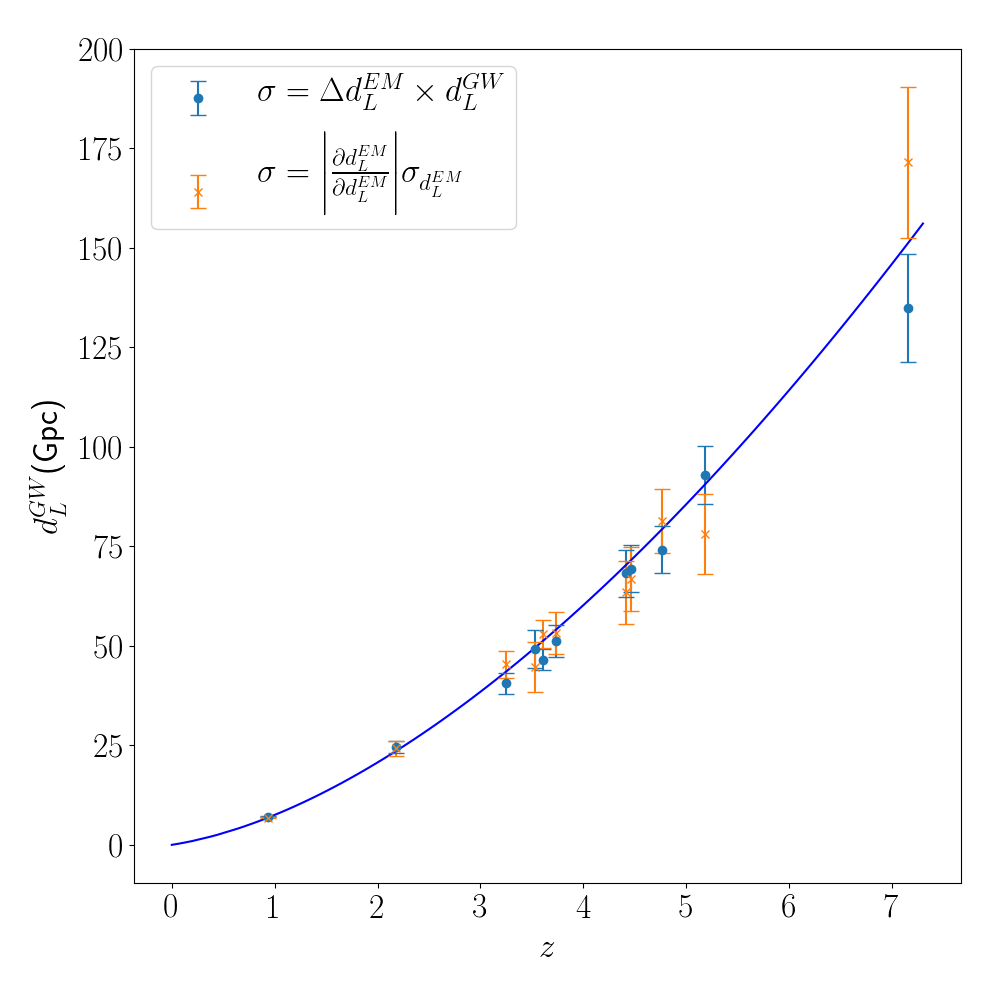}
  \includegraphics[width=0.49\textwidth]{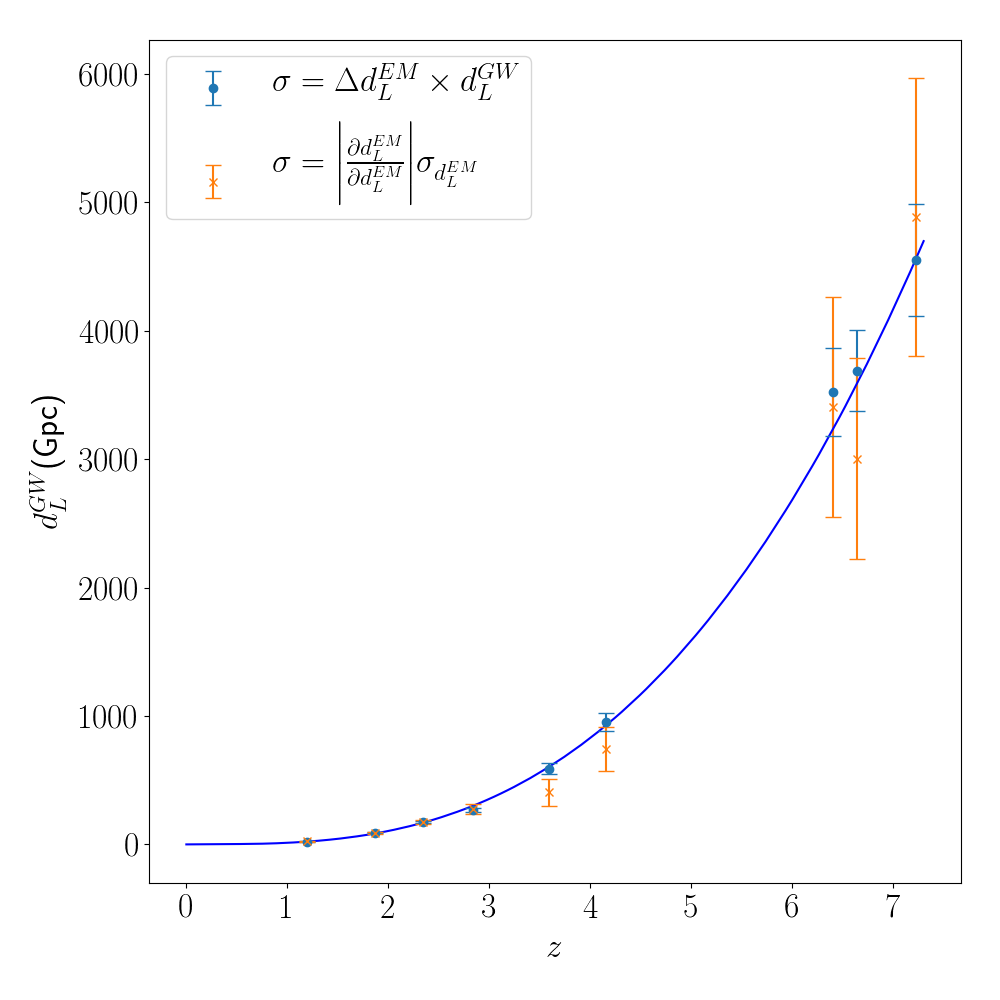}
  \caption{Examples of GW mock data scattered around their true $d_L^{\text{GW}}$ assuming very small (orange) and slightly larger (blue) error bars. \textit{Left}: True cosmological model is characterised by $\theta^{D=5}:(m=1, H_0)$. \textit{Right}: True cosmological model is characterised by $\theta^{D=7}:(m=20, H_0)$.}
 \vspace{-1.em}
\label{fig:sigma}
\end{figure}

Notice that the further away the sources and the more pronounced the modification from GR, i.e. the larger the number of dimensions, the lower the screening scale, and the steeper the transition, the larger the discrepancy between the GW and EM luminosity distance and hence the larger the factor by which the error on $x^{\text{GW}}$ is increased. To understand this, Fig. \ref{fig:sigma} shows simulated GW data scattered around their `true' GW distance for both $\sigma$ stated above for the less `extreme'  $\theta^{D=5}: (n=1, H_0)$ and most extreme $\theta^{D=7}: (n=20, H_0)$ cosmological scenarios.

\bibliographystyle{unsrt}
\bibliography{references}

\end{document}